\documentstyle[eqsecnum,aps,epsf]{revtex}                        

\draft
\preprint{}
\twocolumn

\title{Measurement processes in quantum physics:\\
a new theory of measurements in terms of statistical ensembles}

\author{W. A. Hofer} 
\address{Technische Universit\"at Wien\\
         A--1040 Vienna, Austria}

\begin{document}
\maketitle


\begin{abstract} 
Considering the recently established arbitrariness the Schr\"odinger equation
has to be interpreted as an equation of motion for a statistical ensemble of 
particles. The statistical qualities of individual particles derive from the 
unknown intrinsic energy components, they depend on the physical environment
by way of external potentials. Due to these statistical qualities and wave 
function normalization, non--locality is inherent to the fundamental relations 
of Planck, de Broglie and Schr\"odinger. A local formulation of these 
statements is introduced and briefly assessed, the modified and local 
Schr\"odinger equation is non--linear. Quantum measurements are analyzed 
in detail, the exact interplay between causal and statistical reasons in 
a measurement process can be accounted for. Examples of individual measurement
effects in quantum theory are given, the treatment of diffraction experiments, 
neutron interferences, quantum erasers, the quantum Zeno effect, and 
interaction--free measurements can be described consistent with the 
suggested framework. The paper additionally provides a strictly local and 
deterministic calculation of interactions in a magnetic field. The results 
suggest that quantum theory is a statistical formalism which derives its 
validity in measurements from considering every possible measurement of a 
given system. It can equally be established, that the framework of quantum 
physics is {\em theoretically} incomplete, because a justification of 
ensemble qualities is not provided.
\end{abstract} 


\section{Introduction}

In recent papers we demonstrated that
quantum theory allows for a layer of hidden variables, which determines
the properties of individual electrons and photons 
\cite{HOF95,HOF96A,HOF96B}. The aim of the new theory was to estimate,
whether the current axiomatic interpretation of quantum theory is the
only possible solution to wave--like properties of micro physical
systems, especially since the current standard does not allow for a
{\em physical} interpretation of wave functions. 

By revealing intrinsic potentials inherent to wave properties of
moving particles the wave structures of individual particles could
be deduced, which provided the fundamental relations of quantum 
theory as well as electrodynamics. The Schr\"odinger equation
\cite{SCH26} was identified as a mechanical modification of the 
particle wave equation, and it was demonstrated, that it allows 
for a level of arbitrariness described by the uncertainty relations.

As many discussions with readers of the previous summary of the
theory \cite{HOF96B} showed, the exact range of the concept was not
defined clearly enough. Equally did the logical relation between the
theory and measurements in quantum theory not be clearly determined.
This was partly intended, since the first step of establishing a new
concept must be the definition of fundamental properties and 
relations. As ensembles and interactions of ensembles are somewhat
more complex than single particles, the issue of measurements and
systems of particles was left aside. But as the logical implications of
measurements in quantum theory present a substantial problem not
solved in any other way than by applying the axioms of the
{\em Copenhagen interpretation} \cite{COP26}, the issue surely requires 
to be clarified.

This paper is therefore aimed at determining the {\em exact} interface,
where the deterministic picture of material wave theory -- based on
the qualities of single electrons and photons and their interactions --
is replaced by the statistical picture of quantum theory. It is
equally sought to clarify, how quantum theory combines the two 
different perspectives in a single picture, more precise, what the
mathematical procedures employed actually signify.

In order to clarify the essential issues of this paper some initial
remarks seem necessary, especially in view of current concepts. The
standard framework (classical electrodynamics and quantum field theory) 
employs three central concepts for the description of physical events:
a {\em particle} is thought to be a {\em physical} entity with
definite mechanical qualities (mass and charge, enhanced with the
non--classical quality of spin) and local boundaries 
commonly considered insignificant (this is also the case for de Broglie's
particle and Renninger's conception of the photon).
Dealing with particles, physics tends to consider them as 
reasonably stable units (even in case of short decay times).
A {\em wave} possesses neither limited dimensions, nor can it be 
described in terms of mechanical quantities.  The notion of 
{\em coherence length}, used to describe the maximum
extension of a field in order to produce interference effects, is
already a combination: while in a strictly classical context, the
concept of light pulses with small cross--sections (an everyday
experience in laser--optics), would lead to severe boundary problems
at the local limits of the laser beam, a more practical approach
has to account for the limited extensions of electromagnetic fields
and reduces classical theories to the region well within the
beam boundaries. Given this limited extension of fields in phenomena
of light absorption and emission -- verified in abundance in 
spectroscopic measurements --, it seems evident that the classical
theory cannot be employed to describe events in a large system
without substantial adaptations. Therefore neither the concept of
a mechanical particle (usually considered point--like), nor the
concept of a wave (essentially unlimited) are strictly valid.

These, practical, limitations have so far not found sufficient
resonance in theoretical modeling. More important, these limitations
have not even been used in any fundamental way to modify the physical
frameworks. 

As the Maxwell equations are verified to the point of
certainty, the question of their validity needs to be analyzed from
the viewpoint of limited field extensions. If unlimited space is
thought to be filled by electromagnetic fields, the actual value
of these fields within the coherence length of a beam should yield
results which are not initially compatible with the solutions of the
Maxwell equations, since, after all, energy conservation in the 
classical framework does not account for beam boundaries.
As the solutions of Maxwell's equations are determined but for a
constant of integration, the absolute quantity cannot be
calculated: the solutions therefore do not allow for an evaluation of 
energy density in specific regions of the system. The problems of
compatibility, as soon as quantities are considered, points to
the essential limitations of classical electrodynamics: which is,
in our view, one more reason for the infinity problems inherent 
to quantum electrodynamics. 

Similar considerations apply to the problem of point--like
particles: in this case an aggregation is determined by the absolute
quantity of mass or charge. The energy density then acquires a
statistical meaning. 

The decisive results of previous papers on material wave theory do
not allow for either of these interpretations: as found by considering
the intrinsic properties of {\em single} particles, their internal
structure complies with the Maxwell equation, i.e. with a strictly
deterministic and causal framework. The theory introduced the concept
of a {\em limited extension} of single particles, and it was shown, 
that the local extension does not alter the results achieved. 
The consequence of this notion was developed on a fundamental level. 
On the one hand, the physical units of electrodynamics were analyzed 
in view of interaction processes, achieved by photons: the result of 
this analysis was that the units actually signify {\em mechanical} 
energy densities. On the other hand, it was sought to clarify the 
notion of an {\em energy quantum} in interaction processes: and the 
essential result, in this respect, was that the decisive properties 
are energy densities and not energy quanta.

In this paper, the fundamental ideas of the new approach are developed
in view of current standards. Since a particle, or the limited extension
of a physical entity complying with the Maxwell equations, cannot be
the origin of statistical measurement results, we have to consider an
ensemble of these entities. The structure of the ensemble considered
derives from the, equally derived, fundamental uncertainty in any
evaluation of the Schr\"odinger equation: contrary to Maxwell's 
equations this equation does not provide causal and deterministic
results, because it is essentially arbitrary. The fundamental entity 
of this new approach could be called a {\em wave particle}: wave--like
qualities are retained as internal structures and polarizations of
the intrinsic fields, while the particle aspect is due to its
limited extensions. The duality is therefore replaced by a physically 
rigorous simultaneity: and it  is equally shown, that the classical limit 
is a specific form of the resulting ensemble. An ensemble with exactly 
defined energy values. 

We are well aware that this treatment is not fully causal, because
it includes statistical ensembles. This statistical feature, as will 
be demonstrated, is also inherent to classical electrodynamics: if the 
coherence length of single photons and the intensity of radiation are 
successively decreasing, then the overall formalization of 
electrodynamics {\em must} become a statistical limit. That is not to 
say, though, that the statistical qualities are inherent to the 
interactions or interferences of a single photon: the individual 
process is causal and deterministic, but neither electrodynamics, nor 
quantum theory deal with these single processes, but only with 
statistical ensembles of many single entities.

The main advantage of this new approach as well as its essential
justification is to provide a theoretical framework which is neither
limited to mechanical nor to strictly field theoretical concepts.
The fundamental results of current theories can be derived in a
close to local framework (the only concession to non--locality
is the assumption that frequency and wavelength remain fairly
constant within the considered region), and the concept of
''spin'' as well as the logically puzzling results of diffraction
and interference measurements can be understood without recurring
to extraneous sources or potentials.

In section \ref{qu_en} we develop the principal consequences of
the arbitrariness inherent to the Schr\"odinger equation, by 
including all particle wave functions for a specific energy
eigenvalue $E$. The result is a statistical ensemble, which
is changed in different potential environments. On this basis the
probability amplitude $\psi^{*}(\vec r) \psi(\vec r)$ of the ensemble
is derived, which is a function of $E$ and 
$V(\vec r)$. The wave--particle duality, which
is inherent to conventional models, is found to be a result of
misinterpreting the properties of ensembles as properties of
single particles. In the suggested model the duality is replaced by a
hierarchical structure of single particles, particle ensembles, and
their mathematical description.

Section \ref{ar_nl} proposes a local modification of the 
Schr\"odinger equation. It is shown that the local equation is
non--linear and the consequences for the superposition of two
plane waves are developed: while the wave equation and the
electromagnetic framework are not affected by non--linearity,
the modification leads to interference effects if system
development is evaluated with the Schr\"odinger equation.

In section \ref{meas_co} the collapse of the wave--function due to
external potentials is analyzed, it is shown that the process has
to be seen as change of statistical ensembles.

In section \ref{meas_dif} we analyze the original diffraction experiments
and its modern version, the neutron interferences. Magnetic
interactions are calculated, and it is established that interaction
energy does not depend on the orientation of magnetic fields. 

Section \ref{meas_zeno} and \ref{meas_intf} treat two specific
measurements effects in quantum theory, the quantum Zeno paradox and
the interaction free measurements.  

In section \ref{meas_hv} the statistical model of particle manifolds is
compared to hidden variable theories and to the proofs by von Neumann
and Jauch and Piron, that a statistical interpretation of quantum
theory is contradictory. It is shown, that the particle ensemble,
treated in quantum theory, is not dispersion--free, which renders the
theoretical proofs against hidden variables inapplicable.

As a new theory can only gain by deviating viewpoints and controversial
opinions, any feedback is greatly appreciated: readers are encouraged
to contact me directy via email at the email address:
whofer@eapa04.iap.tuwien.ac.at

\section{Quantum ensembles}\label{qu_en}

Rejecting the orthodox Copenhagen interpretation of quantum theory 
requires, in principle, a clarification of the interface between
causal behavior of the intrinsic properties of single particles
(described by the classical theory of electrodynamics and additional
results on electron--photon interactions), and the results of
measurements in microphysics, which are generally of a statistical 
nature. Or, in David Bohm's words \cite{BOH52}:

{\em The usual interpretation of the quantum theory is self--consistent,
but it involves an assumption that cannot be tested experimentally,
viz., that the most complete possible specification of an individual
system is in terms of a wave function that determines only probable 
results of actual measurement processes. The only way of investigating
the truth of this assumption is by trying to find some other
interpretation of the quantum theory in terms of at present
''hidden'' variables, which in principle determine the precise
behavior of an individual system, but which are in practice averaged
over in measurements of the type that can now be carried out.}

In the previous paper \cite{HOF96B} it was found that the 
Schr\"odinger equation can be seen as the linear differential 
equation of internal particle structures. Its derivation by 
way of the wave equation employed a mechanical concept, the 
total energy of a particle wave. This total energy was, like 
in classical mechanics, determined by kinetic energy of the 
particle and potential energy of a given environment. The 
essential difference from {\em any} current interpretation is the
result, that Schr\"odinger's equation does not completely 
define the intrinsic properties of any single particle, because
it generally omits intrinsic potentials. It could be shown,
in this context, that the immediate consequence of this arbitrariness 
is the fundamental uncertainty in Heisenberg's relations \cite{HEI27}.

In the following section we will determine the structure of
the statistical ensembles pertaining to this fundamental arbitrariness.
To differentiate between solutions applying to single particle
waves from solutions applying to the whole ensemble, we will
use the terms {\em particle}, {\em wave particle}, or 
{\em particle wave} to denote the intrinsic properties of any 
specific particle with finite dimensions, while the term 
{\em ensemble} refers to all particles or particle waves
meeting the described requirement.

\subsection{Quantum ensemble of free particles}

Due to periodic wave functions, the intrinsic potential at an
arbitrary moment $t$ can take any value, and the wave vector
of the problem therefore is not exactly determined, but covers
the whole range from $k^2 = 0$ to $k^2 = \frac{m}{\hbar^2} E_{T}$.
The interpretation of this result reveals a rather interesting
feature of the Schr\"odinger equation. The differential 
equation in its time--free formulation does not only describe
one specific particle or one specific wave function, but a whole
range of individual particle wave functions for every single
point of a given region. When seen from the viewpoint of intrinsic 
particle properties it describes therefore an allowed range of
particle states. And the formalism is basically a statistical
distribution, where every single result has the same statistical
weight. 

Consider a point $\vec r$, where the external potential 
vanishes $V(\vec r) = 0$. Due to the disregard for intrinsic
potentials the Schr\"odinger equation at this location applies
for all particle waves described by:

\begin{eqnarray}\label{ar001}
\vec k^{2}(\vec r) + \vec k_{i}^2(t) = \frac{m}{\hbar^2} E_{T}
\nonumber \\
0 \leq \vec k_{i}^2(t) \leq \frac{m}{\hbar^2} E_{T}
\end{eqnarray}

$ \vec k_{i}^2(t) $ denotes the intrinsic potential, not accounted
for in quantum theory, which is the origin of the inherent
arbitrariness. $E_{T}$ is the total energy of the particle. The
two variables are given by:

\begin{eqnarray}\label{ar002}
E_{T} = m u^2 \qquad 
\frac{\hbar^2}{m}\vec k_{i}^2(t) = \phi_{i}(t) V_{p}
\end{eqnarray}

where $u$ is the velocity and $V_{P}$ the volume of a particle.
The {\em quantum ensemble} is the Fourier integral
over allowed wave states. The wave function $\psi(\vec r)$ can 
then be written as:

\begin{eqnarray}\label{ar003}
\psi(\vec r) &=&  \frac{1}{(2 \pi)^{3/2}} \int_{0}^{k_{0}}
d^3k \, \phi_{0}(\vec k) e^{i \vec k \vec r} \nonumber \\
k_{0} &=& \sqrt{\frac{m}{\hbar^2}E_{T}} 
\end{eqnarray}

Using the Fourier transformation the amplitudes of the ensemble
are consequently:

\begin{eqnarray}\label{ar004}
\phi_{0}(\vec k) = \frac{1}{(2 \pi)^{3/2}} 
\, \int_{-\infty}^{+\infty}
d^3 r\, \psi(\vec r) e^{- i \vec k \vec r}
\end{eqnarray}

The quantum ensemble in simplified accounts of quantum theory
(especially the simplifications in fundamental treatments of wave 
mechanics, see, for example \cite{MES64}) is reduced to its member of 
half the total energy. The reduction follows from the conventional 
solution of the Schr\"odinger equation, where a solution in the 
orthogonal basis of plane waves will be:

\begin{eqnarray}\label{ar005}
\psi(\vec r) &=& \frac{1}{(2 \pi)^{3/2}}\,
e^{i \vec k_{qm} \vec r} \nonumber \\
\phi_{0}(\vec k) &=& \frac{1}{(2 \pi)^3} \int_{-\infty}^{+\infty}
d^3 r\, e^{i (\vec k_{qm} - \vec k) \vec r} = 
\delta^3 (\vec k_{qm} - \vec k) \nonumber \\
\phi_{0}(\vec k_{qm}) &=& 1 \qquad 
\phi_{0}(\vec k \ne \vec k_{qm}) = 0
\end{eqnarray} 

From the viewpoint of material wave theory and equally quantum theory 
the reduction is an approximation (in quantum theory due to particle
properties, while in material wave theory due to the fundamental 
arbitrariness). To describe the full ensemble of possible waves including 
intrinsic potentials the Fourier integral has to be retained, the quantum 
ensemble in this case is:

\begin{eqnarray}\label{ar006}
\psi(\vec r) &=&  \frac{1}{(2 \pi)^{3/2}} \int_{0}^{k_{0}}
d^3k \, \phi_{0}(\vec k) e^{i \vec k \vec r} \nonumber \\
k_{0} &=& \sqrt{\frac{m}{\hbar^2}E_{T}} \quad
E_{T} = m u^2
\end{eqnarray} 

In view of intrinsic properties of particles the
general solution of the Schr\"odinger equation of free particles
considered at a specific location $\vec r$ is thus not a defined 
state of the particle, but a defined range of possible particle
states. The consideration applies to every single point.

The result differs from Bohm's analysis of the measurement process
\cite{BOH66A}. In Bohm's view the Schr\"odinger equation could not
account for an ensemble of particles due to its mathematical
properties. Since it is a linear differential equation of second order
it completely determines the development of the wave function from
initial values. The Schr\"odinger equation, Bohm concluded, is 
therefore the deterministic part of quantum theory. And he continued:

{\em Yet it is not immediately clear how the ensembles, to which
\ldots probabilities refer, are formed and what their individual
elements are. For the very terminology of quantum mechanics contains
an unusual and significant feature, in that what is called the
physical state of an individual quantum mechanical system is 
assumed to manifest itself only in an ensemble of systems.} 

Translating {\em system} into {\em particle}, the very features
Bohm describes as inherent to quantum theory are found by estimating
the effect of arbitrariness on the results provided by the
Schr\"odinger equation. The time--dependent 
Schr\"odinger equation will not be treated separately, because it is 
standard practice in quantum theory to separate the time--dependent 
part either by a linear harmonic including the energy or frequency of 
the particle, or by a unitary transformation yielding the development
of time--free solutions of the Schr\"odinger equation. 

\subsection{Quantum ensemble in external potentials}

An external potential at $\vec r$ basically has two effects:
the range of allowed particle waves and therefore the quantum 
ensemble will be changed, and the internal properties of single
particle waves will be altered. If the potential at $\vec r$ equals 
$V(\vec r)$, the allowed $k$--values will comply with:

\begin{eqnarray}\label{ar007}
k^2(\vec r) &+& k_{i}^2(t) = \frac{m}{\hbar^2} 
\left(E_{T} - V(\vec r) \right) \nonumber \\
0 & \leq & \vec k_{i}^2(t) \leq \frac{m}{\hbar^2} 
\left(E_{T} - V(\vec r) \right) \\
k_{1}^2 &=& \frac{m}{\hbar^2}(E_{T} - V(\vec r)) \quad
E_{T} = m u^2 
\end{eqnarray}

For reasons of consistency the potential $V(\vec r)$ is double
the potential if only kinetic properties are considered. The
range of allowed $k$--values in this case depends on the energy
$E_{T}$ of a single particle as well as the potential applied. 
There are two distinct cases: $E_{T} - V(\vec r)$ is either a
positive or a negative value, corresponding to particle waves
in a potential or to exponential decay of single waves.

\begin{figure}
\epsfxsize=0.9\hsize
\epsfbox{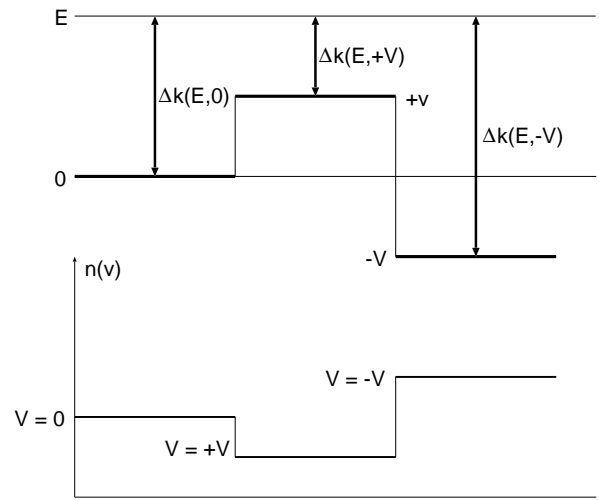}
\vspace{0.5cm}
\caption{Quantum ensemble of particle waves for three 
different potentials. A negative potential increases the number
of allowed $k$--values, a positive potential has the opposite
effect}\label{fig001}
\end{figure}

\subsubsection{Positive solutions $E_{T} - V(\vec r) \ge 0$}

The potential applied can either be a positive or a negative
value, leading to an enhancement or a reduction of the quantum
ensemble of valid solutions. The general solutions for both cases
are then:

\begin{eqnarray}\label{ar008}
\psi(\vec r) &=&  \frac{1}{(2 \pi)^{3/2}} \,\int_{0}^{k_{1}}
d^3k \, \phi_{0}(\vec k) e^{i \vec k \vec r} \nonumber \\
k_{1} &=& \sqrt{\frac{m}{\hbar^2}(E_{T} \pm |V(\vec r)|)} \quad
E_{T} = m u^2
\end{eqnarray}

The range of individual particle waves defines, as in the case of 
a vanishing external potential, an ensemble of particles, which 
comply with the differential Schr\"odinger equation. Any integration 
of the equation therefore also contains a -- hidden -- manifold of
individual particle waves. 

It is interesting to consider the effect of
opposite signs of the external potential. A positive potential essentially
limits the number of individual waves contained in the ensemble, because
it diminishes the range of $k$. A negative potential has the opposite 
effect: the number of waves in the ensemble is increased, and their
statistical weight in the whole system is equally higher.
\mbox{Fig. \ref{fig001}} displays the quantum ensembles for different
external potentials. 

\subsubsection{Negative solutions $E_{T} - V(\vec r) \le 0$}

The mathematical formalism of Schr\"odinger's equation in this case allows
for solutions with a negative square of $\vec k$, equivalent to an
exponential decay of single particle waves:

\begin{eqnarray}\label{ar009}
k^2(\vec r) &+& k_{i}^2(t) = \frac{m}{\hbar^2} 
\left(E_{T} - V(\vec r) \right) \le 0 \nonumber \\
0 & \geq & \vec k_{i}^2(t) \geq \frac{m}{\hbar^2} 
\left(E_{T} - V(\vec r) \right) \\
\psi(\vec r) &=& \frac{1}{(2 \pi)^{3/2}} \,\int_{0}^{k_{1}}
d^3k \, \phi_{0}(\vec k) e^{- \vec k \vec r} \nonumber \\
k_{1} &=& \sqrt{\frac{m}{\hbar^2}(V(\vec r) - E_{T})} \quad
E_{T} = m u^2
\end{eqnarray}

The question in this case concerns not so much the mathematical 
formalism but the physical validity. Considering, that electrodynamics
is a description of intrinsic properties of single particles, applicable
to photons as well as electrons, the results of electrodynamics in different
media should also have relevance for the wave properties of single particles.
And considering, furthermore, that an exponential decay into a medium at a
boundary is one type of solution, the same must generally hold for single
particle waves. Therefore the exponential decay results from wave properties
of single particles and has to be included in the overall picture. It is
a basically {\em classical} solution to boundary value problems and, so far,
no specific feature of a quantum system. 

\subsection{Wave function normalization}

We have not yet defined the amplitude $\phi_{0}(\vec k)$ of single 
components in the quantum ensemble. This can be done by requiring single
particle waves to comply with the mass relations of particles. Since:

\begin{eqnarray}\label{ar010}
\phi^{*}(\vec k) \phi(\vec k') = \frac{1}{(2 \pi)^3}
\phi_{0}(\vec k) \phi_{0}(\vec k') e^{i \vec r (\vec k' - \vec k)}
\end{eqnarray}

an integration over infinite space yields the result:

\begin{eqnarray}\label{ar011}
&&\int_{-\infty}^{+\infty}d^3 r\,\phi^{*}(\vec k) \phi(\vec k') = \nonumber \\
&+& \int_{-\infty}^{+\infty}d^3r \,\delta^3(\vec k - \vec k') 
\phi_{0}(\vec k) \phi_{0}(\vec k') \\
m &=& \int_{-\infty}^{+\infty}d^3 r\,|\phi(\vec k)|^2 =
\phi_{0}^2(\vec k) \nonumber
\end{eqnarray}

Using this amplitude the square of the wave function $\psi(\vec r)$
in different external potentials can be calculated.

\begin{eqnarray}\label{ar012}
&&\int_{-\infty}^{+\infty}d^3r\,|\psi(\vec r)|^2 =
\frac{m}{(2 \pi)^3} \int d^3r
\int_{0}^{k}d^3k d^3k'e^{i \vec r (\vec k' - \vec k)} \nonumber \\
&& =  m \int_{0}^{k}d^3k\,d^3k'
\delta^3(\vec k - \vec k') = \frac{4 \pi m}{3}\,k^3
\end{eqnarray}

And since $k$ can be written in terms of momentum of a particle
the calculation yields the result, that the wave function of the
electron ensemble integrated over the while space is proportional 
to $u^3$:

\begin{eqnarray}\label{ar013}
\int_{-\infty}^{+\infty}d^3r\,|\psi(\vec r)|^2 =
\frac{4 \pi m^4}{3 \hbar^3}\,u^3 = 2.46 \times 10^{-18} \cdot u^3
\end{eqnarray}

The left term,
the integral of the ensemble wave function, does not depend on the
exact location, whereas the right term, the velocity or the wave
vector of the quantum ensemble, clearly is a local function because
it depends on the potential $V(\vec r)$.

\begin{eqnarray}\label{ar014}
\int_{-\infty}^{+\infty}d^3r\,|\psi(\vec r)|^2 &=&
\alpha \cdot u^3(\vec r) \nonumber \\
u(\vec r) &=& \sqrt{\frac{1}{m} (E_{T} - V(\vec r))}
\end{eqnarray}

If $\psi(\vec r)$ is therefore interpreted as the wave function of
any {\em single} particle, the formalism must yield non--local effects.
The result sheds some light on non--locality
in quantum theory. While individual particle waves and the quantum 
ensemble are strictly local, the normalization of $\psi(\vec r)$ and
the subsequent interpretation of $\psi(\vec r)$ as the wave function
of single particles suffices to introduce non--locality into the basically 
statistical framework.

The problem can be avoided by considering intensive properties rather
than quantities. If the potential $V(\vec r)$ is constant throughout
the system, the integral over the wave function $\psi(\vec r)$ can
be written:

\begin{eqnarray}\label{ar015}
\int_{-\infty}^{+\infty}d^3r\,|\psi(\vec r)|^2 =
V_{S} \psi^{*}(\vec r) \psi(\vec r) = m \frac{4 \pi}{3} k^3(\vec r)
\end{eqnarray}

And setting the volume of the system $V_{S}$ equal to a particle
volume $V_{P}$, the right term reduces to density $\bar \rho$
and a function of energy:

\begin{eqnarray}\label{ar016}
\psi^{*}(\vec r) \psi(\vec r) &=& \bar \rho \,\frac{4 \pi}{3} 
\left(\frac{m}{\hbar^2}\right)^{3/2}
\,\left(E_{T} - V(\vec r)\right)^{3/2}
\end{eqnarray}

The effect of these definitions will be that the 
{\em probability density} of the quantum ensemble can be
determined, it will be a local function and depend on the potential
as well as particle energy, while the average density 
$\bar\rho$ of the ensemble, which is a non--local function, 
can be calculated by an integral over the whole system of varying
potentials:

\begin{eqnarray}\label{ar017}
&&\int d^3r \,|\psi(\vec r)|^2 := 1 \quad
\bar\rho = \frac{1}{\int d^3r \,\left[
\left(E_{T} - V(\vec r)\right) \right]^{3/2}}\nonumber \\
&&\psi^{*}(\vec r) \psi(\vec r) =
\frac{\left(E_{T} - V(\vec r)\right)^{3/2}}{\int d^3r \,\left(
E_{T} - V(\vec r)\right)^{3/2}}
\end{eqnarray}

It is evident that these definitions yield a local and steady 
function for the probability density of the quantum ensemble. The
relation between probability density and energy at a point $\vec r$
is non--linear. As the probability density of the quantum ensemble
must be positive definite, this solution does not include situations
where the total energy is lower than the external potential $V(\vec r)$.

\subsection{Boundary conditions}

The decisive difference in the interpretation of quantum systems
is the statistical interpretation of the manifold contributing to
solutions to Schr\"odinger's equation. The difference becomes 
especially obvious, if boundary conditions are considered.
One of the easiest examples in quantum theory, which suffices
for this purpose, is the square potential well. We take a one
dimensional potential well, the external potentials described by:

\begin{eqnarray} \label{ar102}
V &=& 0  \quad \forall \,|x| \le x_{0} \nonumber\\
V &=& V_{0}  \quad \forall \,|x| \ge x_{0} 
\end{eqnarray}

To solve the problem we have two consider the behavior of single
members of the quantum ensemble as well as the behavior of the whole
ensemble. The limiting $k$ values can again be inferred from the
solution of the one--dimensional Schr\"odinger equation, they will be
for $E_{T} < V_{0}$:

\begin{eqnarray} \label{ar103}
k_{0}^2 &\le& \frac{m}{\hbar^2} E_{T} \quad  |x| \le x_{0} \nonumber \\
k_{0}'^2 &\le& \frac{m}{\hbar^2} (V_{0} - E_{T}) \quad  |x| \ge x_{0}
\end{eqnarray}

On the physical level any single member of the quantum ensemble then
is subject to boundary conditions at the two boundaries $ \pm x_{0}$.
The problem on this level has to be treated by field theory, and we
will use an analogy in electrodynamics. As electromagnetic waves, and
also electron waves, are described by the same basic equations, the
results of electrodynamics on wave reflection and exponential decay
into a region, where oscillation is physically impossible, must
equally apply. And as the energy is not sufficient for wave propagation
in the potential barrier, the result in this area should be an 
exponential decay. As we generally are confronted with incident, 
reflected and penetrating waves, the three components of the wave
are a wave of positive and a wave of negative propagation in the
region $ |x| < x_{0}$, and a decaying component in the region 
$|x| > x_{0}$. It has to be noted, that the concept of a particle
does not arise at this level. If we use plane waves, it can be 
assumed that we only use the real or the complex component to satisfy
the boundary conditions, since this component is periodic it will be
proportional to a component of the intrinsic electromagnetic fields,
and the boundary conditions imposed are then basically electromagnetic
conditions. 

Considering individual members of the ensemble, the lowest $k$ value
should correspond to maximum decay in the potential, while the member
with maximum total energy should display maximum penetration. The
relation between an arbitrary wave vector $k_{1}$ and its corresponding
member $k_{2}$ must therefore be:

\begin{eqnarray}\label{ar104}
k_{2}^2 = \frac{m}{\hbar^2} V_{0} - k_{1}^2 \qquad
k_{1}^2 + k_{2}^2 = \frac{m}{\hbar^2} V_{0}
\end{eqnarray}

The wave functions in the three separate regions shall be described
by standard solutions.
Accounting for  the boundary conditions for steady transition of
the particle wave the coefficients can be determined and the
solution for an individual wave is therefore, equivalent to the
solution in quantum theory \cite{COH77}:

\begin{eqnarray}\label{ar105}
&&  \phi_{0} \cdot e^{k_{2} x} \qquad \qquad \qquad
x \le - x_{0} \nonumber \\
\phi(x) &=& \phi_{0} \cdot 
e^{- k_{2} x_{0}} \frac{\cos k_{1} x}{\cos k_{1} x_{0}}
\qquad |x| \le x_{0} \nonumber\\
&&  \phi_{0} \cdot e^{- k_{2} x} \qquad \qquad \qquad
x \ge x_{0}
\end{eqnarray}

The normalization conditions for a member of the ensemble allow to
calculate the amplitude of the particle wave:

\begin{eqnarray}\label{ar106}
\phi_{0}(k_{1},k_{2}) = 
\sqrt{\frac{m k_{2}}{1 + k_{2} x_{0}}}e^{k_{2}x_{0}}
\cos k_{1} x_{0}
\end{eqnarray}

And the total ensemble can equally be calculated by integrating
over the full range of allowed $k$ values.

\begin{eqnarray}\label{ar107}
&&\int_{\infty}^{\infty} dx |\psi(x)|^2 = 2 \int_{0}^{\infty} dx \left[
\theta(x_{0}-x) \int_{0}^{k_{0}} dk_{1}\, \phi_{0}^2(k_{1}) \times \right.
\nonumber \\
&& \left. \times e^{- 2 k_{1} x_{0}} \frac{\cos^2(k_{1} x)}
{\cos^2(k_{1} x_{0})} 
+ \theta(x-x_{0}) \int_{0}^{k_{0}'} dk_{2}\, 
\phi_{0}^2(k_{2}) e^{- 2 k_{2} x} \right]
\nonumber
\end{eqnarray}

The amplitude $\phi_{0}(k)$ must finally be renormalized, and the
square of the wave function then describes the probability distribution
of the whole ensemble. The procedure described is equal to the standard
procedure in quantum theory, since a single electron in quantum theory 
must be described as a Fourier integral over $k$ space. The essential
difference is, though, that in the present context $k$ space is not
unlimited, the cutoff is determined by the energy $E_{T}$.

The parallel treatment of quantum ensembles in different environments
and the conventional treatment in quantum theory could basically be
performed for arbitrary environments and initial conditions, for the
present purpose it suffices to demonstrate, that the framework of
quantum theory allows for a statistical interpretation. The causal
level of the concept treats single particle waves and their boundary
conditions, while the statistical level refers to manifolds of
possible physical properties. Quantum theory is therefore a statistical
concept, and the statistical distribution is equal to the quantum 
ensemble defined in this section.

\subsection{Local ensembles}

The particle waves treated in the context of intrinsic properties
were {\em local} entities, the volume of any single wave is therefore
finite. The normalization of the wave function for a full quantum
ensemble of allowed energy values and arbitrary locations within the
system has the effect, that the results of any integration of 
Schr\"odinger's equation are no longer strictly local. Consider
a system with an arbitrary number of individual particle waves,
which could be named, in an allusion to Louis de Broglie's term
\cite{LDB27}, {\em solitons}. In de Broglie's view a soliton
possesses insignificant volume, its motion within
the limits of the wave function was later referred to a hidden 
quantum potential (Bohm \cite{BOH52}) or to Brownian motion 
(Nelson \cite{NEL66,NEL67}). 
In our view the term soliton shall only signify the exact energy 
value and the limited local extensions of a single particle wave, 
retaining all its intrinsic properties described by a field 
theoretical approach. The difference of the current approach to
de Broglie's concept is the interpretation and the significance
of the wavefunction for the qualities of ensembles: while de
Broglie's suggestion that the wavefunction be associated with an
ensemble of identical particles is accepted (these particles possess
different positions within a considered system), the influence of 
the wavefunction on motion of the particles (or the concept of
a pilot--wave) is rejected. In our view the wavefunction is not a
strictly physical entity in ensembles, but a statistical measure.
The current theory therefore combines two hitherto separate methods:
the statistical interpretation of quantum theory by Ballentine
\cite{BAL70}, and the causal interpretation of the theory by
de Broglie and Bohm \cite{LDB26,LDB27B,BOH52}. But while the 
de Broglie--Bohm theory is highly non--local \cite{HOL93},
the current framework
retains locality and refers physical processes to a change of the
internal structure of the particles. That this essentially local
concept is equivalent to the conventional one in the description of
measurement processes, will be demonstrated in the following sections.
But that it additionally allows to treat fundamental processes in
a local and determinist manner, seems to be a major achievement.

Let us take an ensemble of particles with exactly defined 
energies. We will describe, how the quantum ensemble can be
reduced to constant energy values in the following
sections, currently we only assume, that it is feasible. Then
a soliton covers only a comparatively small region of the
whole system $S$, which we shall not specify. In this case the quantum
ensemble consists of an arbitrary number of solitons, covering the
whole of the considered system. The only arbitrariness retained 
is the local arbitrariness, the reason being again, that the
time variable in Schr\"odinger's equation is accounted for by a
unitary transformation removing any explicit local dependency. 
The solutions of Schr\"odinger's equations then cover the whole 
local range of the system. Since the amplitude of the wave function 
is arbitrary, the exact number of solitons cannot be estimated. 
The local ensemble $\psi_{\nu}(\vec r)$ is given by:

\begin{eqnarray}\label{ar108}
\left\{\psi(\vec r,t \in [-\infty,+\infty])\right\} \rightarrow
\left\{\psi_{\nu}(\vec r)\right\} \nonumber \\
\vec r \in S \qquad \nu \in [0,\infty]
\end{eqnarray}

And together with the normalization condition for the square of
the wave function we are forced to accept the probability 
interpretation of the wave function.

\begin{eqnarray}\label{ar109}
\int_{\nu} \int_{S} d^3r \, d\nu \, 
|\psi_{\nu}(\vec r)|^2 =: 1 \nonumber \\
\quad \Rightarrow \quad
\int_{\nu} d\nu \,
|\psi_{\nu}(\vec r)|^2 = w(\vec r)
\end{eqnarray}

The concept clearly does not allow for any causal interpretation of
the wave function. The structure of physical statements is therefore
multidimensional. The causal dimension reflects interactions of
single solitons due to fundamental processes, which can be seen as 
photon interactions. The interaction processes were extensively
analyzed, and it was shown that the subsequent quantization rules
apply to every single point of the intrinsic particle structure
\cite{HOF96B}. But the causal dimension is not sufficient, neither
to account for the features of quantum systems, nor indeed for 
a physical formalization of measurement processes, which generally
have to account for statistical ensembles. On this level the probability
interpretation as well as the subsequent Copenhagen interpretation
of quantum theory are {\em practical conventions} to deal with the 
statistical nature of microphysical processes. It must be stressed,
though, that it is not a physical interpretation deriving from the
structure of microphysical processes, but only an interpretation based 
on the features of quantum ensembles.

The local ensemble is a special case of the full 
quantum ensemble, the case where the energy of individual particles 
is exactly determined. Comparing classical field theories with 
ensemble structures, it will be seen further on, that the local
ensemble is also the classical limit of quantum theory.

The ensemble is introducing, by way of normalization
procedures, a non--local component into fundamental statements
of quantum theory. That is, though, not the whole story. Based on
this result it could be assumed, that some clever structure of
physical statements will enable us to refer the statements to 
a strictly local theory. That this is not possible, even on the
level of fundamental particles without fundamental changes in the
mathematical structure of quantum theory, can be demonstrated as 
follows:

Take a particle (photon or electron) of finite volume $V_{P}$
and exactly defined energy $\hbar \omega$. Consider now the
frequency or wavelength of the particle wave at an arbitrary 
position $\vec r$ within the particle. Then the frequency $\omega$
at this point is given by:

\begin{eqnarray} \label{ar110}
\omega(\vec r) = \frac{\rho(\vec r) \cdot V_{P}}{\hbar} \cdot 
u(\vec r)^2 
\end{eqnarray}

Since the volume of the particle cannot be described as a function
of location (or only, again, via a normalization condition 
$\rho(\vec r) \cdot V_{P} = m_{P}$), the frequency of the wave is
not a function of location: the wave itself therefore has a non--local
element to its structure, which remains only insignificant, as long
as the volume is not measurable. 

\subsection{Classical limit}

The intrinsic functions of a single particle waves describe the
classical limit of physical theories, the case when the energy of
the particle is exactly defined. Since the wave function (due to the
relation $\rho \propto \psi^2$ see \cite{HOF96B}) has the same
periodicity as intrinsic electromagnetic fields, the classical
electromagnetic solutions for an infinite system must be proportional
to the wave functions for an infinite local ensemble. The
classical limit of quantum theory therefore is the local quantum
ensemble.

\subsection{Spreading of a wave packet}\label{meas_spread}

It may seem that the definition of the ensembles is only justified
in view of intrinsic potentials, that it is not originally a concept
of quantum theory. That this is not the case can be shown by a
calculation of developments with the time--dependent Schr\"odinger
equation. The notorious {\em spreading of a wave packet} due to the
application of Schr\"odinger's equation cannot be interpreted in
a physical way, because the implicit experimental result has always
been falsified. 

If it is not a physical effect, it is nevertheless a
feature inherent to the qualities of the wave function. And since the
wave function is, in conventional reasoning, a measure for the
probability density of particles, the problem remains, why the
statistical nature of a quantum system of free particles is altered.
This effect, so far not really understood, is a common source of
irritation, and concepts have been put forth to eliminate it in a
modified version of quantum theory (see, for example, Mackinnon 
\cite{MAK78,DAT85}). 
  
\begin{figure}
\epsfxsize=1.0\hsize
\epsfbox{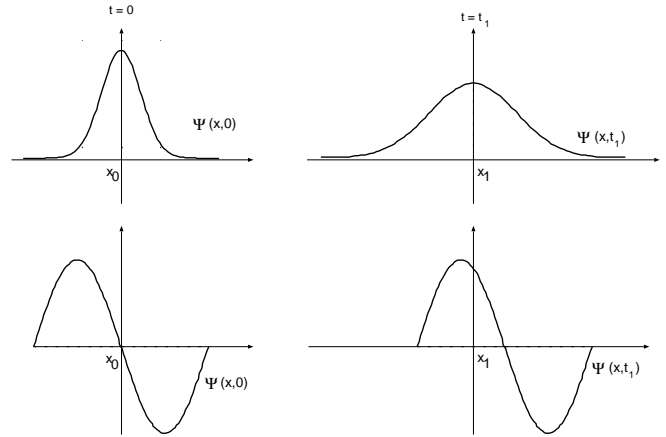}
\vspace{0.5cm}
\caption{Development of ensembles. An ensemble of
arbitrary energy and Gaussian distribution will develop into
a quantum ensemble of equal probability for all energy
values (top). The development of a particle with exactly defined
energy due to the Schr\"odinger equation or the wave equation leads
to a local ensemble over the whole local range of the system
(bottom)}\label{fig002}
\end{figure}

We consider a quantum system of presently undefined $k$--vectors
and local extension. Then the quantum ensemble at an initial moment
$t = 0$ shall be given by the Fourier components of the wave function,
$ \hat\psi(\vec k)$, the development of initial components shall be
described by plane waves, for the sake of simplicity we treat the
one--dimensional model:

\begin{eqnarray}\label{ar201}
\psi(x, t) = \int dk \,e^{i(kx - \omega t)} \hat\psi(k)
\end{eqnarray}

The essential feature to consider is the development of an initial
distribution of a system only governed by the Schr\"odinger
equation. The two initial distributions we estimate are the local
ensemble with exactly defined energies and wave vectors and an
inhomogeneous quantum ensemble, where the wave function shall be
centered around a value $x_{0} = 0$:

\begin{eqnarray}\label{ar202}
\psi_{1}(x, t = 0) &=& e^{- \frac{x^2}{b^2} + i k_{0}x} \nonumber \\
\psi_{2}(x, t = 0) &=& e^{i k_{0} x}
\end{eqnarray}

The Fourier transform can be calculated from the initial state
of the wave functions, it will be:

\begin{eqnarray}\label{ar203}
\hat\psi_{1}(k) &=& \int dx \,\psi_{1}(x,0)\,e^{- i k x} =
e^{-(k - k_{0})^2\,b^2/2} \nonumber \\
\hat\psi_{2}(k) &=& \int dx \,\psi_{2}(x,0)\,e^{- i k x} =
\delta (k - k_{0})
\end{eqnarray}

And the wave functions of the two systems are at $t > 0$:

\begin{eqnarray}\label{ar204}
\psi_{1}(x, t > 0) &=& \frac{e^X}{\sqrt{b^2 + \frac{i \hbar t}{m}}} 
\nonumber \\ 
X &=& \frac{1}{2} \left(b^4 + \frac{\hbar^2 t^2}{m^2}\right)^{-1} \cdot
\left( - k_{0}^2 b^2 \frac{\hbar^2 t^2}{m^2} + \cdots \right) \nonumber \\
\psi_{2}(x, t > 0) &=& \exp \left(i (k_{0} x - \omega_{0}t)\right)
\end{eqnarray}

The norm of the two wave functions is consequently:

\begin{eqnarray}\label{ar205}
&&\left| \psi_{1}(x, t > 0) \right|^2 =
\left( 1 + \frac{\hbar^2t^2}{m^2 b^4}\right)^{-1} \times \nonumber \\
& \exp & \left[ - b^{-2} \left( 1 + \frac{\hbar^2t^2}{m^2 b^4}\right)^{-2}
\cdot \left( x - \frac{\hbar k_{0}}{m} t \right)^2 \right] \nonumber \\
&&\left| \psi_{2}(x, t > 0) \right|^2 = 1
\end{eqnarray}

The norm of the wave functions in the infinite limit 
$ t \rightarrow \pm \infty $, the probability density, will
in both cases be an equal distribution over the whole range of the
system: in case of the initial Gaussian distribution it does no longer
reflect the initial structure of the Fourier transform, different 
components can then no longer be distinguished. It is, in the infinite
limit, an equal distribution over the whole range of $k$ values, or
an infinite quantum ensemble. The second wave function will retain its
kinetic properties, although it will cover the whole local range of
the system. The wave therefore has the same qualities as the local
ensemble. The development of the two ensembles is displayed in
Fig. \ref{fig002}

As it is always possible to decompose an arbitrary distribution of
initial $k$ values in Gaussian distributions, the result holds quite
generally. In case the system is left to itself, an application of the 
Schr\"odinger equation thus leads to a gradual change of the system
qualities until, in the infinite limit, the qualities are equal to
a quantum ensemble.

The spreading of the wave packet therefore appears to be a statistical
process of restructuring: it reveals an entropy--like principle inherent
to the conventional formulations of quantum theory, which is not quite
easy to analyze. Apart from the requirement, that the final state of
a system must be an equal distribution of physical qualities, it cannot
be based on any physical foundation. That the equal distribution is 
physically relevant in measurement processes, will be derived later
on. But that this groundstate of physical properties is arrived at
without any interaction seems hard to accept.

There are two possibilities to account for this feature: (1)
Either the restructuring is referred to some potential not covered
by field theories, in this case we would have to recur to Bohm's
quantum potential \cite{BOH52}. (2) The initial conditions contain
an assumption which affects the physical properties of particles.
The answer to this problem can be found on the level of intrinsic
properties of particles. Since 
$\psi(x)^2 \propto \rho(x) \propto \phi(x)$, an inhomogeneous 
distribution like $ \psi_{1}(x, 0)$ gives rise to an intrinsic
potential $\phi(x)$, described by:

\begin{eqnarray}\label{ar206}
\psi_{1}(x,0) &=& \psi_{0} e^{i k_{0} x} \qquad
\psi_{0} = e^{- x^2/2b^2} \nonumber \\
\phi(x, 0) &=& \psi_{0}^2 = e^{- x^2/b^2}  
\end{eqnarray}

And the system is therefore not, as implied by the mathematical
formulations, free of forces, but will experience a force along the
direction $x$:

\begin{eqnarray}\label{ar207}
F_{x} = - \frac{\partial \phi}{\partial x} = 
\frac{2 x}{b^2} e^{- x^2/b^2}
\end{eqnarray}

The exact evaluation of this existing force and its effects for
the propagating wave need not occupy us at this point: it suffices
to prove that the initial assumption (free waves) cannot be sustained
in view of intrinsic properties, and that the result of quantum theory,
which served as a proof against Schr\"odinger's initial concept of
physical waves, is physically invalid.  

Along this line of reasoning we may reconsider the question of
quantum ensembles from the viewpoint of intrinsic potentials and
forces. The most general form of a wave function is given by its
Fourier integral over all components:

\begin{eqnarray}\label{ar208}
\psi(\vec r) &=& \int d^3k \, \psi_{0, \vec k}(\vec r, \vec k)
e^{i \vec k \vec r}
\end{eqnarray}

The potentials due to the qualities of the amplitude are then
responsible for intrinsic potentials additional to the constant
potential of a plane wave. They are given by:

\begin{eqnarray}\label{ar209}
\phi_{i}(\vec r, \vec k) =
u^2 \, \left|\psi_{0, \vec k}(\vec r, \vec k)\right|^2 =
\frac{\hbar^2 k^2}{m^2} \left|\psi_{0, \vec k}(\vec r, \vec k)\right|^2
\end{eqnarray}

The forces within the propagating wave are either periodic -- the
total potential of the plane wave is constant --, or they are forces 
due to the properties of the amplitude. These forces will be:

\begin{eqnarray}\label{ar210}
\vec F = \frac{\hbar^2 k^2}{m^2} \left[
\psi_{0, \vec k}^{*}\, \nabla 
\psi_{0, \vec k} + 
\psi_{0, \vec k} \, \nabla
\psi_{0, \vec k}^{*} \right]
\end{eqnarray}

A stable state of the system can only be expected, if these
forces vanish. The equilibrium condition for a system of particles
described as plane waves is therefore:

\begin{eqnarray}
\psi_{0, \vec k}^{*} \, \nabla 
\psi_{0, \vec k} + 
\psi_{0, \vec k} \, \nabla
\psi_{0, \vec k}^{*} = 0
\end{eqnarray}

The two ensembles defined, the local ensemble as well as the
general quantum ensemble, comply with this condition since in
both cases the amplitudes $ \psi_{0, \vec k}(\vec k) $
do not depend on $\vec r$. The distribution used for the calculation
of the spreading wave packet clearly is not compatible with this
condition. 

However, given the condition of stability, the quantum ensemble 
seems unnecessarily restrictive: it is not excluded,
that the amplitudes change with the wave vector $\vec k$, although
the quantum ensemble does not allow for an interpretation in terms
of inhomogeneous $\vec k$ distributions. The restriction is not
a physical result, meaning that it does not derive from physical
processes, but a logical consequence of the fundamental arbitrariness
of Schr\"odinger's equation (see the discussion).

\section{A local Schr\"odinger equation}\label{ar_nl}

As the conventional framework is non--local, it might be interesting
to consider its local modification. A feasible method to regain locality
is a different basic set of variables in field theory.
If the concept of {\em particles} is given up altogether, and if
the decisive variables of physical systems are the intrinsic values
of density $\rho$, charge density $\sigma$, as well as the intrinsic
fields of motion $\phi_{k}$ and electromagnetic complements $\phi_{E}$,
and their correlating vector fields $\vec p$, $\vec E$, and $\vec B$,
the Planck and de Broglie relations can be formulated in a local
way. We take the result on hydrogen atoms, where the atomic radius
equals $3.3 \times 10^{-10}$ m \cite{HOF97B}. And we assume that the 
density of electron mass at the atomic radius shall be the density of
a free electron. The volume $V_{el}$ of a free electron shall therefore
be:

\[
V_{el}^0 = 2 \pi R_{0}^3 \approx 2.26 \times 10^{-28} m^3
\]

Therefore the Planck and de Broglie relations can be formulated as local
statements, the frequency and wavelength of a free electron given
by the relations:

\begin{eqnarray} \label{ar111}
\lambda(\vec r, t) \cdot u(\vec r,t) &=& 
\frac{2 \pi \beta_{el}}{\rho_{el}^0(\vec r,t)}
\nonumber \\
\rho_{el}^0(\vec r,t) \cdot \vec u^2(\vec r,t) &=& 
\beta_{el} \, \omega(\vec r,t) \\
\beta_{el} = \frac{\hbar}{V_{el}} &\approx& 4.67 \times 10^{-7} [kg/m s]
\nonumber
\end{eqnarray}

In view of current results, the volume of an electron is far too high.
But as recently established \cite{HOF96B}, the energy exchange in 
interactions not only applies to the electron as a whole, but to
every single point within the electron. Scattering experiments are
therefore no way to detect definite electron extensions. 

Rewriting the Schr\"odinger equation for electrons in a strictly
local manner, we get furthermore (density $\rho_{0}$, frequency
$\omega$, and velocity $\vec u$ equal to the functions defined in 
(\ref{ar111}), indices and variables omitted for brevity):

\begin{eqnarray}\label{ar112}
\rho_{0} \left(
- \frac{\beta_{el}^2}{\rho_{0}^2} \, \triangle
+ \phi(\vec r, t) \right) \psi = 
\beta_{el} \omega\, \psi \nonumber \\
\psi(\vec r,t) = \psi_{0} \exp i \left(
\frac{\rho_{0} \, \vec u}{\beta_{el}} \vec r - \omega t \right)
\end{eqnarray}

It can be seen, that the local version of Schr\"odinger's equation
is no longer linear, the linearity of the relation, which corresponds
to the principle of superposition only holds in the non--local case.

The general case of a local Schr\"odinger equation, in its time--free
and time--dependent form can be written as the nonlinear system of the 
following equations:

\begin{eqnarray}\label{ar113}
A \qquad & \left[
- \beta_{el}^2\, \triangle +  \rho_{0}^2\, \phi(\vec r, t) \right] \psi = 
\rho_{0} \beta_{el} \omega \, \psi \nonumber \\
B \qquad & \left[
- \beta_{el}^2 \, \triangle + \rho_{0}^2\, \phi(\vec r, t) \right] \psi = 
i \rho_{0} \,\beta_{el} \, \dot{\psi}\\
C \qquad &\rho_{0} \, = |\psi(\vec r,t)|^2 \nonumber
\end{eqnarray}

Or if we define the Hamiltonian density ${\cal H}(\vec r,t)$, 
the local system of equations for electron motion will be given by:

\begin{eqnarray}\label{ar114}
{\cal H}(\vec r,t) &:=& \rho_{0} \left(
 - \frac{\beta_{el}^2}{\rho_{0}^2} \, \triangle
+ \phi(\vec r, t)\right) \nonumber \\
{\cal H} \, \psi &=& \beta_{el} \omega \, \psi \\
{\cal H} \, \psi &=& 
- \frac{\beta_{el}}{i} \, \frac{\partial \psi}{\partial t} 
\nonumber \\
\rho &=& \psi^{*} \psi \nonumber
\end{eqnarray}

In view of these results it can be concluded that non--locality is
a feature inherent to the conventional formulation of quantum theory,
it can be derived from its fundamental statements without any 
experimental consideration. The local formulation of these statements
yields non--linear differential equations, a local theory therefore
will not comply with the principle of superposition.

If we calculate the effect of superimposing two single solutions
of the local Schr\"odinger equation, we may proceed from the two
plane waves at $\vec r,t$:

\begin{eqnarray}\label{ar115}
\psi_{1} = \psi_{0}^{1} \exp i \left(
\frac{\rho_{0}^1 \, \vec u_{1}}{\beta_{el}} \vec r - \omega_{1} t \right)
\quad |\psi_{0}^1|^2\left(u_{1}^2 + \phi\right) = \beta_{el} \omega_{1}
\nonumber \\
\psi_{2} = \psi_{0}^{2} \exp i \left(
\frac{\rho_{0}^2 \, \vec u}{\beta_{el}} \vec r - \omega_{2} t \right)
\quad |\psi_{0}^2|^2\left(u_{2}^2 + \phi\right) = \beta_{el} \omega_{2}
\nonumber
\end{eqnarray}

A superposition of the two solutions leads to a relation containing
additional interference--terms of the two original waves:

\begin{eqnarray}\label{ar116}
\psi_{s} &:=& \psi_{1} + \psi_{2} \\ \nonumber \\
{\cal H} \psi_{s} &=& \beta_{el}\, \frac{|\psi_{0}^1|^2 \omega_{1} +
|\psi_{0}^2|^2 \omega_{2}}{|\psi_{1} + \psi_{2}|^2} + \nonumber \\
&+& \phi \,\frac{|\psi_{1} + \psi_{2}|^4 - |\psi_{0}^1|^4 - |\psi_{0}^2|^4}
{|\psi_{1} + \psi_{2}|^2} \ne \beta_{el}(\omega_{1} + \omega_{2})
\nonumber
\end{eqnarray}

The fundamental relations are therefore essentially non--linear
and depend on the interference of the two original waves.
The difference between the non--local and the local formulation
of the equation, furthermore, implies the intensity of the potential
in interactions of electrons. If the amplitudes of the original
waves $\psi_{1}$ and $\psi_{2}$ are unity, then a superposition of
the two waves to $\psi_{s}$ means, that the potential, and therefore the
photons of emission, must be:

\begin{eqnarray}\label{ar117}
\phi &:=& \phi_{1} + \phi_{2} =
\omega_{1} \cdot \Gamma(\vec r, t) + \omega_{2} \Gamma(\vec r, t) 
\nonumber \\
\Gamma(\vec r, t) & \equiv & \beta_{el}
\cdot \frac{1 + 2 \cos \triangle \varphi}
{4(1 + \cos \triangle \varphi)^2 - 2} \\
\triangle \varphi &=& \frac{\vec u_{1} - \vec u_{2}}{\beta_{el}} \vec r
- (\omega_{1} - \omega_{2})t \nonumber
\end{eqnarray}

The effect must be considered a result of the mechanical 
formulation, which does not allow for any simple evaluation of
particle interactions. The wave equation is not affected in 
the same way: it remains a linear differential equation of second 
order still allowing for the superposition of single solutions, 
as can easily be established:

\begin{eqnarray}\label{ar118}
\left( \triangle - \frac{1}{u^2} \frac{\partial^2}{\partial t^2}
\right)\psi = 0 \nonumber \\
\psi = \psi_{0} \exp i\left(\frac{\rho_{0} \vec u}{\beta} \,\vec r - 
\omega t \right)
\end{eqnarray}

And the dispersion relation for the monochromatic wave yields
the local form of Planck's relation or the relation between the
total energy density and frequency of the plane wave.

\begin{eqnarray}\label{ar119}
\beta = \frac{\rho_{0} \vec u^2}{\omega} \qquad
\vec k = \frac{\rho_{0} \vec u}{\beta}
\end{eqnarray}

Since electromagnetic formulations are derived from the wave 
equation, the theory of classical electrodynamics is equally not 
affected.

\section{Collapse of the wave function}\label{meas_co}

In his rather fundamental and comprehensive analysis of
measurement processes in quantum theory Ballentine \cite{BAL70,BAL84}
proceeded from two mutually exclusive statements on the quality
of the state concept, i.e. that (1) {\em a pure state provides a
complete and exhaustive description of an individual system}, and
(2) {\em a pure (or mixed) state describes the statistical properties
of an ensemble of similarly prepared systems}. The subsequent analysis
of measurement processes proved that \cite{BAL84} {\em any interpretation
of the type (1) \ldots is untenable}. Ballentine remarks, furthermore,
that the {\em collapse of the wave function} is a necessary 
consequence of an interpretation type (1) and, since this interpretation
of the wave function is disproved, the theory gives no evidence
of this collapse to occur.

The process is conventionally described as follows: a particle with
its wave function $\psi(q)$, where $q$ denotes some generalized
coordinates is thought to interact with a measurement apparatus
described by the wave function $\phi(\xi)$, where $\xi$ equally
describes the generalized coordinates of the measurement system.
After the interaction the combined wave function of the measurement
apparatus and the particle is described by 
$\psi_{n}(q)\,\phi_{n}(\xi)$, where $n$ is a single eigenstate and
pointer position. In von Neumann's view (see Ref. \cite{NEU32}) 
the process must occur in two steps: (1) the interaction of particle
and measurement device leads to a superposition of states described 
by:

\[
\Psi = \sum_{n} \psi_{n}(q)\,\phi_{n}(\xi)
\]

(2) The observation of the measurement leads to the collapse of the
wave function to its actually measured state given by:

\[
\Psi = \psi_{n}(q)\,\phi_{n}(\xi)
\]

The formalization is not changed if instead of discrete pointer
positions a range of continuous measurement values is considered.
One just has to substitute the index $n$ by a continuous index
$\nu$ describing the range of measurements. The basic problem is 
that this process cannot be described within the framework of
quantum theory, which leads, in the more orthodox interpretations,
to the negation of every process which is not an actual measurement
result (Heisenberg's view in Ref \cite{REN53}), or even to
metaphysical interpretations of the effects of consciousness (which
seemed to be Bohr's view, stating that the process of observation
is essentially {\em non-causal} and {\em irreversible}).
In more ingenious interpretations the result has been been used as
the basis of a {\em many world interpretation} (Everett \cite{EVE57}),
where every outcome of a measurement occurs in a different universe.
For a complete survey as well as a critical analysis of the
many world interpretations see Adrian Kent's original paper
\cite{KEN90}, or his recent update in the eprint archive \cite{KEN97}.
Recently theoretical models focus on the concept of 
{\em quantum--entanglement}. Clearly, the problem is still rather
disputed. 

If the state vector of a system, or the wave function in quantum
theory were an exhaustive information about the system, the logical
problems seem indeed severe if not unsurmountable. The situation
changes, though, if one concedes that the quantum mechanical description
of systems is essentially arbitrary and does not provide a full
account of physical variables, as demonstrated in the previous 
paper \cite{HOF96B}. The results obtained do not exclude a collapse of the
wave function during a measurement process leading to a {\em coherent
superposition} of different measurement values (or, in Ballentine's
words "pointer positions"), although the specific wave function is not a
complete and exhaustive description of a quantum system: this is,
incidentally, the result obtained in the previous section. 
From a physical point of view, it remains to clarify, how the measurement
affects the quantum ensemble, and if the total number of ensemble 
members is reduced in this process.

A simple example of the reduction of the wave function in a
measurement process is a retarding field analyzer frequently
employed in LEED (low energy electron diffraction) measurements.
The reason that we do not use the commonly used spin--measurements
is the oscillating feature of particle--spins when defined
according to the definition in quantum theory \cite{HOF96B}.

\begin{figure}
\epsfxsize=1.0\hsize
\epsfbox{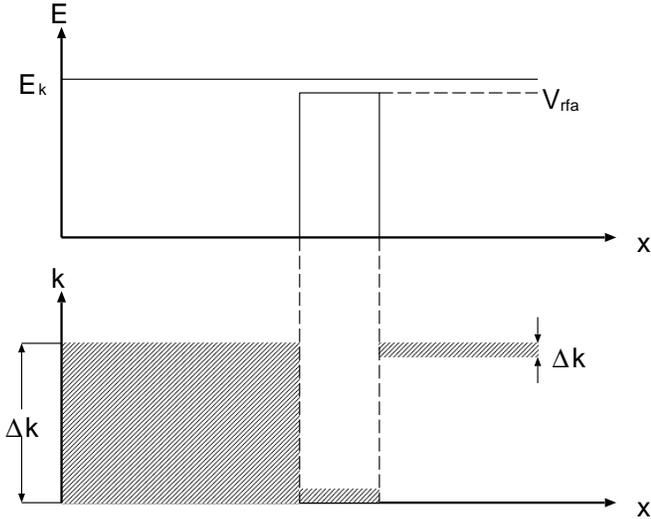}
\vspace{0.5cm}
\caption{Reduction of the wave function due to retarding field
analyzer. The positive potential $V_{rfa}$ is lower than the 
total energy of the ensemble limit $E_{k}$ (top), the selection
of members of the ensemble leads to a reduction of the statistical
ensemble after the measurement (bottom)
}\label{fig003}
\end{figure}

A retarding field analyzer is basically a positive potential, 
assumed rectangular for simplicity, which selects only electrons
above a certain energy threshold. We equally assume, that the
electrons initially are free, their energy shall be given by
a value $E_{k}$ (see Fig. \ref{fig003}). 

From a strictly causal point of view, the electrons below an exactly
defined threshold value $E_{rfa} < E_{k}$ cannot pass the filter and the
number of electrons after the filter is therefore reduced to single
particles with an energy above the threshold value. Naively one could
assume, that the wave function after the barrier must be the same
as before the barrier, since all the electrons had sufficient energy.
This assumption is only correct, though, if intrinsic potentials remain
unconsidered.

As the calculation in quantum theory accounts for the arbitrariness
by integrating over all possible particle states at a given location
$\vec r$, and since the range of allowed $k$--values depends on the
level of kinetic energy, the total of density $\rho(\vec r)$ at an
arbitrary location before the analyzer will be:

\begin{eqnarray}\label{if011}
\rho(\vec r) = \psi^{*}(\vec r) \psi(\vec r) =
\int_{0}^{k_{0}}d^3 k \phi^{*}(\vec k,\vec r) 
\phi(\vec k,\vec r) \nonumber \\
k_{0}^2 = \frac{2 m}{\hbar^2}\,E_{k}
\end{eqnarray} 

while after the analyzer the wave function will be limited to states
with energy values higher than the threshold:

\begin{eqnarray}\label{if012}
\rho'(\vec r) = \psi'^{*}(\vec r) \psi'(\vec r) =
\int_{k_{1}}^{k_{0}}d^3 k \phi^{*}(\vec k,\vec r) 
\phi(\vec k,\vec r) \nonumber \\
k_{1}^2 = \frac{2 m}{\hbar^2}\,(E_{k} - E_{rfa})
\end{eqnarray}

Clearly the wave function, or the quantum ensemble, have been reduced. 
The interesting question now seems to be whether this collapse is a causal 
process, a statistical one, or a process only found in quantum mechanical 
systems. 

Let's say we want to make sure, that only particles of a specific 
energy will be part of the measurement process. Thus some sort of
preselection has to be established, which guarantees that the
total potential of the particles meets the required conditions.
This sort of preselection requires, though, that the particles 
are initially {\em prepared}, the same way the retarding field
analyzer makes sure, that only particles above the threshold
will pass. If the particles are prepared, then the reduction will
occur at the preparation process, exactly in the same way as
above. Then all the particles prepared will pass the analyzer
and no further reduction is required. If particles are not 
prepared, then the result is the previous one, although this result is
a consequence of a lack of knowledge about individual particles.
It is therefore, like in statistical kinetics, a result of
statistical distributions of intrinsic potentials at every 
arbitrary point $\vec r$ of the system, formalized in quantum
theory by way of the fundamental arbitrariness. 

The measurement of particle energies by retarding fields is
therefore a causal and deterministic process on a statistical
ensemble. The same applies to the reduction of the wave function.
A special quantum process, apart from the intrinsic 
features of single particles, is therefore not required.      

In an ideal limit, where the field analyzer selects only particles
of a defined energy $E_{k}$, the quantum ensemble after the measurement
is reduced to the local ensemble. In this case the wave function 
describes an ensemble of exactly defined energies and infinite
extensions: this ensemble has the structure of infinite electromagnetic
fields, it is therefore equal to the classical limit of quantum
theory. The same could basically be achieved by a monochromator,
and, as will be seen further down, particle properties in this case
only prevail as limited regions of space of individual particle
waves. That this limitation is not described by classical formulations,
is equally understandable, since the classical limit is based on an
infinite local ensemble.

\section{Diffraction experiments}\label{meas_dif}

\subsection{Double--slit experiments}

Diffraction experiments have long been difficult to analyze since
the exact interplay between causal reasons and statistical considerations
remained something of a puzzlement to logical reasoning. Especially
the existence of interference terms in the notorious double--slit
experiments has been hard to explain (see for example Bohm's analysis
of the double slit experiment in \cite{BOH66A}, or recent experiments
on complementarity and quantum erasers \cite{HER95}). The main problem, from
the viewpoint of logical reasoning, is the fact that the path of a
single particle through one aperture (an electron for example) should
be determined by the second aperture, since it is evident, from the
results of measurements, that the interference pattern changes with
the state of the second aperture. There remained basically only two
solutions to the problem: (1) the Copenhagen interpretation 
\cite{COP26} stating
that an individual particle is only to be seen as a statistical
ensemble of particles, the statistical quality is therefore inherent
to the trajectory of any single one (which is essentially the complementarity
argument), or (2) Bohm's interpretation
that the trajectory of a single particle is exactly determined by the
Schr\"odinger equation, and that the statistical qualities derive from
hidden variables \cite{BOH66A}.

From the viewpoint of intrinsic particle properties and the interpretation
of Schr\"odinger's equation as an equation of motion for statistical
manifolds of particles, both solutions are partly correct, although
neither of them provides a full account of the problem.  

The Copenhagen interpretation is correct, as far as the quantum 
ensemble of the measurement is concerned. That the non--local formulation
of quantum theory is theoretically incapable of describing the
field type interactions of a {\em single} particle wave in the potentials
of the slit environment, cannot be used -- as the dogmatic version of
the interpretation holds -- to assume any metaphysical connection between
the observed particle and the observing experimenter. It is, in the present
context, the result of the local quantum ensemble, which yields the
characteristic interference patterns. As every single particle wave only
covers a small local range of the whole ensemble, the {\em individual}
measurement result exhibits point like features. And since this
specific ensemble is equivalent to the classical limit, a classical theory
of diffraction will provide a suitable model to describe the experimental
result.

The question of self interference of a particle with a volume similar to the 
volume of an electron derived in section \ref{ar_nl} thus indicates the point,
where the particle interpretation breaks down: all the intrinsic relations 
remain valid, if an electron beam is thought to be split in two components
(which in fact is the only possible interpretation concerning neutron 
interferences treated further down), the same holds for every single point
in interactions. That the measurement after the double--slit system then
is registering a limited region of space for the trajectory, is only a
consequence of the limited volume of space covered by the wave structure
of a ''single'' electron. In this case as well as in the case of neutron 
interferences the problem to a consistent physical interpretation of events 
(a causal chain at that), is the particle picture of quantum theory 
deriving from the basically mechanical outlook and the subsequent 
normalization.

Self interaction of single particles is therefore an unsuitable term:
the internal structure of the particle interacts with the slit environment 
(this is the causal and local level, which is {\em not} described by quantum 
theory, but classical field theories).
And as the ensemble of particles in this measurement covers the whole local
range, all results of the classical limit will eventually be covered 
(this is the statistical level, described by the ensemble wave function
as well as the classical variable of intensity).
The interface between the causal level and the statistical one in this
case is the local range of intrinsic structures covering, eventually,
the whole system.

Bohm's result is not quite consistent with the qualities of the
Schr\"odinger equation: as the Schr\"odinger equation is not a causal
-- i.e. physical -- description of single particles (due to non--locality
in the normalization and arbitrariness because of omitted intrinsic 
potentials), it can only describe the behavior of ensembles, while the
proposed deterministic quantum potential \cite{BOH52} implies physical 
origins for the change of the wave function during measurement. This 
interpretation, in our view, combines two aspects of the problem which 
have nothing in common: the wave function is not a {\em physically} 
valid description of any single particle, therefore the change of the 
wave function has a physical (field interactions) as well as a statistical 
(over the whole local range of the system) aspect. Reducing the physical
aspect to the statistical one (which is, essentially, the content of
the Copenhagen interpretation), or reducing the statistical aspect to
physical ones (which is the content of Bohm's theory) is equally
unjustified: the whole picture is only emerging, if both aspects are
included simultaneously.

To display the essential features of these experiments, it suffices to
describe diffractions by a modification of classical electrodynamics.
The theoretical account sheds some light on the seeming paradox
of single particles and individual measurements, which are, in the
limit of infinite repetitions, equal to classical interference patterns
\cite{TON89}. Let the wave function of a particle (photon or electron)
in the vacuum be described by the Helmholtz equation:

\begin{eqnarray}
\left( \triangle + k^2\right) \psi(\vec r) = 0
\end{eqnarray}

With Green's theorem and using a vacuum Green's function 
we may rewrite the conditions for $\psi(\vec r)$ to:

\begin{eqnarray}
\psi(\vec r) &=& \oint_{R(V)} d^2 \vec f' \,
\frac{e^{i k |\vec r - \vec r'|}}{4 \pi |\vec r - \vec r'|} \times
\\ & \times & \left[ \nabla' \psi(\vec r') +
\frac{2 \pi i \psi(\vec r')}{\lambda}
\frac{\vec r - \vec r'}{|\vec r - \vec r'|}
\left(1 + \frac{i}{k |\vec r - \vec r'|}\right)  \right]
\nonumber 
\end{eqnarray}

The surface $R(V)$ denotes the boundary of the volume of integration.
To account for particle separation we describe the wavefunction
of the ensemble as an integral over all measurements:

\[
\psi(\vec r) = \int\limits_{- \infty}^{+ \infty}
dt \psi_{t}(\vec r,t)
\]

The time dependency and the limited extensions of an individual
particle are included with a $\delta$ function, the wave function
at the aperture of the diffractometer is then:

\begin{eqnarray}
\psi(\vec r') =  \int\limits_{- \infty}^{+ \infty}
dt \, \delta^3(\vec r' - \vec c_{p} t) \cdot
e^{ik |\vec R - \vec c_{p} t|}
\end{eqnarray}

$- \vec R $ shall be the location of the particle source and $c_{p}$
the velocity of the particle. If we neglect the derivatives of the
$\delta$ function (which, after all, shall only denote the existence
of single entities), the following relation is derived:

\begin{eqnarray}
\psi_{t=0}(\vec r) &=& e^{ik R} \Gamma(k,\vec r) \\
\Gamma(k,\vec r) &=& \frac{i}{2 \lambda} \oint_{R(V)} d^2 \vec f'
\frac{\vec r - \vec r'}{|\vec r - \vec r'|^2}
\left(1 + \frac{i}{k |\vec r - \vec r'|}\right)
e^{i k |\vec r - \vec r'|} \nonumber
\end{eqnarray}

With a Kirchoff approximation the calculation yields the familiar
results of classical electrodynamics: the amplitude $\Gamma(k, \vec r)$
depends on the geometrical setup of the diffractometer and particle
wavelength. But due to the $\delta$ function, it yields these
amplitudes for {\em individual} particles passing the device. 

This, rather simple, example reveals much of the logical problems
of quantum theory in interference measurements. If the concept of a
particle is retained, then the wavefunction at the moment $t = 0$, the
moment when it passes the slit, contains already the future
measurement results in form of the amplitude $\Gamma(\vec k, \vec r)$.
But since the amplitude depends on the number of slits -- which,
for a single particle, cannot be a determinant of particle motion through
a specific one -- the experimental result (interference fringes depending
on the experimental setup) cannot be described by a causal {\em and} 
local model. A causal theory therefore {\em must be} non--local 
(which applies to the de Broglie--Bohm approach). If the particle
model is given up and replaced by a model of single wave--particles
with dimensions comparable or larger than the slit geometry, the
amplitude given by classical calculations results essentially from 
the requirements of continuity after the interaction (contained in the
form of the Green's function), while interference is referred to 
a statistical effect due to varying local distributions. It seems to
confirm the interpretation of classical electrodynamics as a theory,
which gains its validity from two separate origins: the formalization
of internal particle structures and the consideration of infinite local
ensembles.

The total wavefunction of the system $\psi(\vec r)$ and the wavefunction
of single measurements $\psi_{t}(\vec r)$ are numerically equal but for the
delta function. While the total wavefunction, therefore, may be
said to contain the infinite repetition of the measurement in the
form of measurement {\em results}, the individual wavefunction
contains the same information in the form of measurement 
{\em probabilities}. The wavefunction in this case has a double 
significance: a statistical as well as a determinist one.

\subsection{Neutron interference}

As a further example we may consider the modern version of these 
measurements, the neutron interference experiments by
Rauch and Zeilinger \cite{RAU92,ZEI81,GREN84}. 

We postulate initially that neutrons are physical entities similar to
electrons or photons, in that they possess wave like properties described
by a wave function $\psi$ of single particles, intrinsic potentials to
account for periodic mass distributions, and they shall be subject to 
the fundamental Planck and de Broglie relations in non--local formulation.
Since the theoretical framework so far has not been extended to nuclear
fields and properties, the postulate is a generalization yet without
an indisputable proof. Additionally, we take intrinsic magnetic fields
into account by postulating, that neutron mass in motion possesses an
intrinsic magnetic field of a specific orientation $\vartheta$, which
shall be perpendicular to the axis of particle motion $\vec u$.
 
\begin{figure}
\epsfxsize=1.0\hsize
\epsfbox{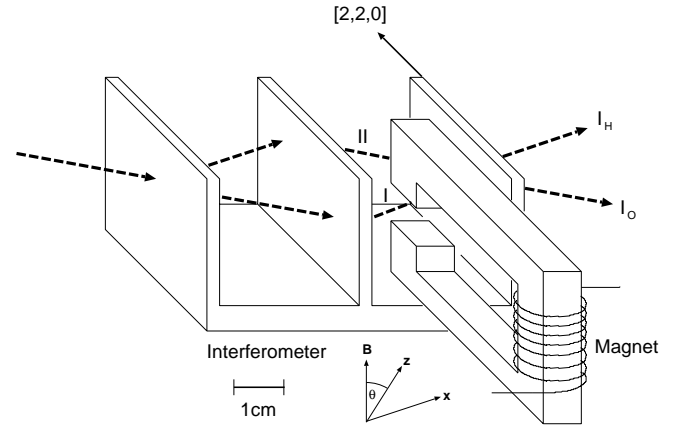}
\vspace{0.5cm}
\caption{Neutron interferometer for spin superposition measurement
according to Rauch. The incident neutron beam is split 
coherently at the first plate and reflected at the middle plate. The 
two beam components $I$ and $II$ are coherently superposed at the third 
plate of the interferometer, a static magnetic field in the path of 
beam $I$ leads to interference patterns depending on magnetic 
intensity}\label{fig004}
\end{figure}

The most interesting results were obtained by a perfect crystal
interferometer based on amplitude division, which yields a macroscopic 
(in the range of cm) beam separation. The intensity of the incident
neutron beam was in every case such that only a single neutron passed
the interferometer at a given moment. The separated beams could be
manipulated by an insertion of attenuation and absorption devices,
in single but highly significant experiments magnetic fields were
employed to change the spin--orientation \cite{RAU92}. According to
Rauch all experimental results can be accounted for, if they are
evaluated in the wave picture of quantum theory, while the particle
picture can only be preserved by applying the concept of de Broglie's
pilot waves (see Fig. \ref{fig004}).

From the viewpoint of material waves the result is not surprising:
as the incident neutrons were initially passing a monochromator, the
resulting quantum ensemble is the local one, which is, as 
established in section \ref{qu_en}, equal to the classical limit of
quantum theory. It does seem, therefore, completely acceptable to
treat the experiments within the framework of classical optics. And
the theoretical calculations then are equal to the classical ones
for x--ray interferences, where the amplitude of the wave function
and electromagnetic intensity are related by ($I_{qt}$ is the
intensity due to quantum theory, $I_{em}$ the intensity obtained
in classical electrodynamics):

\begin{eqnarray}\label{if013}
&I_{qt}& \propto \psi^{*} \psi \propto |\vec E|^2 \propto I_{em} 
\nonumber \\ &\Rightarrow& I_{qt} \propto I_{em}
\end{eqnarray}

There are, though, two results which require a closer scrutiny:
the first is the proximity of experiments with single neutrons to
the thought--experiments by Renninger \cite{REN60} on single photons, 
the second is the change of wave properties by external magnetic 
fields.  

As derived in the section on photons of the previous paper \cite{HOF96B}
the mass of a photon cannot be confined to an infinitesimal volume
within the electromagnetic field, if Maxwell's relations are correct
descriptions of the intrinsic electromagnetic fields. The kinetic
properties of photons thus had to apply to every single point
of the electromagnetic field, and the total energy density within the
photon therefore was an intrinsic constant. The experimental and
theoretical basis of this result was the deterministic relation between
the electromagnetic fields of a photon and the experimental result.
If the experimental results of neutron interferences are, therefore,
described by the deterministic theory of classical electrodynamics,
the same must apply to the relation between kinetic and electromagnetic
properties of neutrons. And the kinetic properties of a neutron shall
then prevail throughout the region of its complementary intrinsic
fields: the initial assumptions about neutrons are then backed by
the experimental results of interference measurements in conjunction
with their theoretical description. This result suggests a wide range
of speculative questions about the {\em physical} nature of neutrons,
which cannot, at present, be treated, since the qualities of nuclear
fields and interactions have not yet been included into the framework
of material wave theory.

\subsection{Magnetic interactions of particles}

Quantum theory bases the interaction of magnetic fields and particles
on the intrinsic qualities of spin and magnetic momenta. Spin is a
non--local property, in previous calculations it was found, that the
concept is theoretically questionable \cite{HOF96B}. The following
calculation is a {\em local} and {\em deterministic} deduction of 
magnetic interactions, based on intrinsic electromagnetic fields
and the field equations of electromagnetic properties as well as
intrinsic potentials:

\begin{eqnarray}\label{if015}
\frac{1}{u^2} \frac{\partial \, \vec E}{\partial t} &=&
\nabla \times \vec B \qquad
- \frac{\partial \, \vec B}{\partial t} =
\nabla \times \vec E \nonumber \\
\phi_{em} &=& \frac{1}{2} 
\left( \frac{1}{u^2} \vec E^2 + \vec B^2\right)
\end{eqnarray}

We consider the change of intrinsic fields due to a constant external
magnetic field $\vec B_{ext}$, the field vectors shall be given by:

\begin{eqnarray}\label{if016}
\vec B &=& (0, 0, B_{0}) \cos(k_{0} x - \omega_{0} t) \nonumber \\
\vec E &=& (0, E_{0}, 0) \cos(k_{0} x - \omega_{0} t) \\
\vec B_{ext} &=& (0, - \sin \vartheta, \cos \vartheta) B_{ext}
\nonumber
\end{eqnarray}

Accounting for the dynamic qualities of the process by linear 
increase of the magnetic field $t \in [0,\tau]$, the internal
fields will be at $\tau$:

\begin{eqnarray}\label{if017}
E_{y}' &=& E_{0} \cos(k_{0} x - \omega_{0} t) - 
B_{ext} \frac{\cos \vartheta}{\tau} x \nonumber \\
E_{z}' &=& - B_{ext} \frac{\sin \vartheta}{\tau} x \\
B_{y}' &=& - B_{ext} \sin \vartheta \nonumber \\
B_{z}' &=& B_{0} \cos(k_{0} x - \omega_{0} t)
+ B_{ext} \cos \vartheta \nonumber
\end{eqnarray}

Additional informations about the system can be inferred from the
relation between the variables $x$ and $t$ as well as from the
relation between amplitudes from Eq. \ref{if015}:

\begin{eqnarray}\label{if018}
\frac{x}{\tau} = u_{0} \quad E_{0} = u_{0} B_{0}
\end{eqnarray}

The electromagnetic potential due to interaction with the
magnetic field is then given by:

\begin{eqnarray}
2 \phi_{em} &=& \left[B_{0} \cos(k_{0}x - \omega_{0}t) - 
B_{ext} \cos \vartheta\right]^2 + \left[B_{ext}\sin \vartheta\right]^2
+ \nonumber \\
&+& \left[B_{0} \cos(k_{0}x - \omega_{0}t) + 
B_{ext} \cos \vartheta\right]^2 + \left[B_{ext}\sin \vartheta\right]^2
\nonumber 
\end{eqnarray}
\begin{eqnarray}\label{if019}
\phi_{em} &=& B_{0}^2 \cos^2(k_{0}x - \omega_{0}t) + B_{ext}^2 
\end{eqnarray}

The result is interesting due to two features: (1) The potential
of interaction does not depend on the angle $\vartheta$ of the
magnetic field: it can therefore not be formalized as the scalar
product of an intrinsic magnetic moment $\vec \mu$ and an external
field $\vec B_{ext}$, or only, if the magnetic moment is a 
non--local variable: the non--local definition of particle spin
in quantum theory can therefore be seen as a different expression
of an equivalent result. And the motivation for this definition has
to be seen in the missing account of deterministic and dynamic
developments of the intrinsic variables. 

The result confirms a
conclusion already drawn by analyzing electron photon interactions:
the framework of quantum theory is essentially limited to interactions,
its logical implications only become obvious, if interaction
processes are considered. In the context of particle spin it explains, why
spin in quantum theory {\em cannot} be a local property: 
{\em because} interactions do not depend on the direction of field
polarization. 

(2) The electromagnetic
potential of the particle is higher than the original potential.
This result leaves two possibilities: either total energy density
of the particle remains constant -- which should be the case for 
neutral particles, which do not alter their energy due to 
magnetic interactions --, or the kinetic energy density of the 
particle is equally altered by interactions: which should apply
for charged particles. In both cases the kinetic potential during
magnetic interaction is changed, the alteration can be described by:

\begin{figure}
\epsfxsize=1.0\hsize
\epsfbox{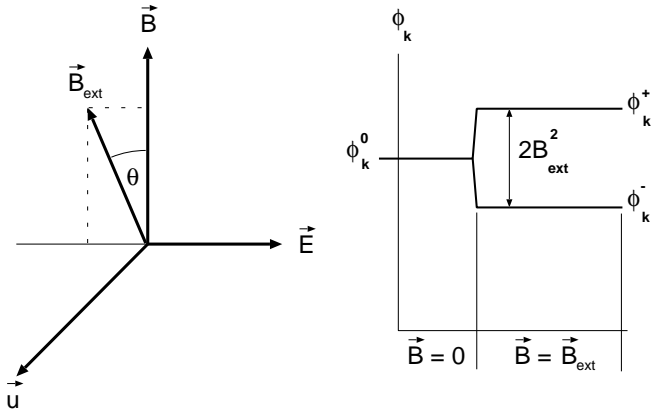}
\vspace{0.5cm}
\caption{Intrinsic electromagnetic fields and applied magnetostatic
field of the neutron beam (left). Shift of kinetic potential due
to interactions (right)}\label{fig005}
\end{figure}

\begin{eqnarray}\label{if020}
\phi_{k}' = \phi_{k}^0 \pm B_{ext}^2
\end{eqnarray} 

Fig. \ref{fig005} displays the intrinsic electromagnetic fields and the 
energy terms due to magnetic interactions.
Due to the relations between kinetic potential and density of
mass $\phi_{k} = \rho u^2$ and the relation between the wave function
and density of mass $\rho \propto \psi^2$ the properties of the
wave function in the region of interaction will equally be changed,
which means, that posterior superposition of the two separated beam
parts will yield a changed interference pattern. The easiest way to
calculate the changes in the affected beam is by estimating the 
difference of velocity. Since:

\begin{eqnarray} \label{if021}
\langle \phi_{k}' - \phi_{k}^0\rangle =: \triangle \phi_{k} =
- \bar \rho \,(\triangle u)^2 = - B_{ext}^2 
\end{eqnarray}

$\bar \rho$ denotes denotes average density of the beam, as the
wave length is much shorter than the macroscopic region of the
magnetic field in the interaction process, averaging is physically
justified. Then the phase difference $\alpha$ of the beam after 
$t_{1} = l/u_{0}$ seconds, where $l$ is the linear dimension of the
magnet, will be:

\begin{eqnarray}\label{if022}
\alpha &=& 2 \pi \,\frac{\triangle u \cdot t_{1}}{\lambda} 
= 2 \pi \left(\frac{l}{\lambda}
\cdot \frac{B_{ext}}{\sqrt{\bar\rho \,u_{0}^2}} -  n \right) \nonumber \\ 
n &\in& N
\end{eqnarray}

\begin{figure}
\epsfxsize=1.0\hsize
\epsfbox{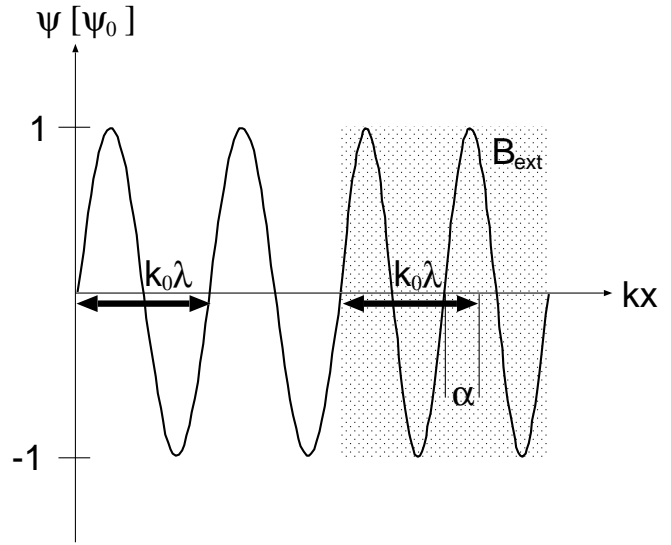}
\vspace{0.5cm}
\caption{Phase shift $\alpha$ in a magnetic field. The phase shift 
of the beam component $I$ depends on the scale of the external 
magnetic field $B_{ext}$}\label{fig006}
\end{figure}

The theoretical result is consistent with the experimental result
by Rauch, that the phase of the beam is linear with the intensity
of the magnetic field applied \cite{RAU92}. Fig \ref{fig006} displays
the phase shift of the beam component $I$ in a magnetic field.

That this phase shift is sufficient for an experimental proof of the 
$4 \pi$--symmetry of spinors seems to be a matter of convention, since
it depends, essentially, on the scaling of the magnetic fields in terms
of kinetic potentials. All that can be inferred from measurements
is that magnetic fields affect the phase of the neutron beam, and
equally, that this effect does not depend on the orientation of the
magnetic field or the incident beam: both results can in a local
and deterministic manner be accounted for by this deduction of
magnetic interactions. 

It should be noted, that the theoretical concept is only applicable
to the local ensemble or the classical limit of quantum theory
(As a consequence of monochromatic neutron beams).
If the ensemble considered contains particles of arbitrary energies
then an equivalent theoretical framework additionally has to account
for the phase shifts of specific members of the ensemble: a close
to classical interference pattern in this case cannot be expected.

\subsection{The quantum eraser}

That the intrinsic properties and, especially, the polarization
of the intrinsic fields is decisive for interference measurements
can be demonstrated by an analysis of {\em quantum eraser} phenomena.
In this case the which--path information of the photon is said to
preclude interference. Conventionally, the measurement is formalized
as follows \cite{SCU82}:

The amplitude of an incident photon of horizontal polarization 
is split coherently in two separate beams, described by the quantum 
state vector ($1$ denotes the first, $2$ the second of the two paths)

\begin{eqnarray}
\psi_{12}^0  = \frac{1}{\sqrt{2}} \left(
\psi_{1,H} + \psi_{2,H} \right)
\end{eqnarray}

The square of $\psi$, or the probability density in this case contains
an interference term $\psi_{1,H}^{*} \psi_{2,H}^{ }$:

\[
|\psi_{12}^0|^2 = \frac{1}{2} \left(
|\psi_{1,H}|^2 + |\psi_{2,H}|^2 + \psi_{1,H}^{*} \psi_{2,H} +
\psi_{2,H}^{*} \psi_{1,H} \right)
\]

If a polarization rotator changing the polarization of the beam
 to vertical (V) orientation is placed in path 1, the interference 
pattern is no longer observable, and the measurement yields random
results for the local probability density on the measurement screen.
In quantum theory the result is referred to the orthogonality of the
two states $H$, and $V$, and the state vector of the photon described
by

\begin{eqnarray}
\psi_{12}^1 &=& \frac{1}{\sqrt{2}} \left(
\psi_{1,V} + \psi_{2,H} \right) \\
|\psi_{12}^1|^2 &=& \frac{1}{2} \left(
|\psi_{1}|^2 + |\psi_{2}|^2 \right) \nonumber
\end{eqnarray}

The result can be changed by inserting a diagonal polarizer into
the path of the recombined beams, in this case the wave function and
probability density will again show interference effects: the 
which--path information, connected to the polarization of the two
separate beams is said to have been ''erased''.

\begin{eqnarray}
\psi_{12}^2 &=& \frac{1}{2 \sqrt{2}} \left(
\psi_{1} + \psi_{2} \right)\left( \psi_{V} + \psi_{H}\right) \\
|\psi_{12}^2|^2 &=& \frac{1}{4} \left(
|\psi_{1}|^2 + |\psi_{2}|^2 + 2Re \left[ 
\psi_{1}^{*} \psi_{2}\right]\right) \nonumber
\end{eqnarray}

The result can be understood in the context of intrinsic properties
and polarizations of intrinsic fields, since the intensity of 
electromagnetic radiation is described by the electromagnetic 
potential $\phi_{em}$ (for a general description the field
vectors are assumed complex):

\begin{eqnarray}
\phi_{em} = \frac{1}{2} \left(\
\frac{1}{c^2} |\vec E|^2 + |\vec B|^2 \right)
\end{eqnarray}

If the beam of horizontal polarization (direction $x$) is split
into two separate beams, the electric and magnetic fields after
recombination will be:

\begin{eqnarray}
\vec E_{12}^0 = \frac{1}{\sqrt{2}} \left(
E_{1} \vec e_{x} + E_{2} \vec e_{x} \right) \nonumber \\
\vec B_{12}^0 = \frac{1}{\sqrt{2}} \left(
B_{1} \vec e_{y} + B_{2} \vec e_{y} \right)
\end{eqnarray}

where the fields $E_{2}, B_{2}$ contain the phase information 
$e^{i \varphi}$. The intensity measured after recombination will 
consequently contain interference terms:

\begin{eqnarray}
\phi_{em}^0 = \frac{1}{2} \left(
\frac{1}{c^2} |E_{1}|^2 + |B_{1}|^2 \right) 
\left(1 + cos \varphi \right)
\end{eqnarray}

A polarization rotator in path 1 changes the polarization of
the electric and magnetic fields to $\vec e_{y}$ and $- \vec e_{x}$
respectively, and the intensity after recombination is then not
affected by the phase $\varphi$:

\begin{eqnarray}
\vec E_{12}^1 = \frac{1}{\sqrt{2}} \left(
E_{1} \vec e_{y} + E_{2} \vec e_{x} \right) \nonumber \\
\vec B_{12}^1 = \frac{1}{\sqrt{2}} \left(
- B_{1} \vec e_{x} + B_{2} \vec e_{y} \right)
\end{eqnarray} 

\begin{eqnarray}
\phi_{em}^1 = \left(
\frac{1}{c^2} |E_{1}|^2 + |B_{1}|^2 \right)
\end{eqnarray}

If the recombined beam is passing through a diagonal polarizer
(plane of polarization in $xy$--direction), the electromagnetic
fields after polarization are:

\begin{eqnarray}
\vec E_{12}^2 = \frac{1}{2} \left(
E_{1} \vec e_{xy} + E_{2} \vec e_{xy} \right) \nonumber \\
\vec B_{12}^2 = \frac{1}{2} \left(
- B_{1} \vec e_{yx} + B_{2} \vec e_{yx} \right)
\end{eqnarray}

And the intensity of the beam shows again the interference pattern
of the phase $\varphi$:

\begin{eqnarray}
\phi_{em}^2 = \frac{1}{4} \left(
\frac{1}{c^2} |E_{1}|^2 + |B_{1}|^2 \right) 
\left(1 + cos \varphi \right)
\end{eqnarray}

Mathematically, the description by way of intrinsic potentials and
polarizations yields the same result as the conventional calculation
in quantum theory. However, the interesting aspect of the effect is its
interpretation. While in the conventional framework the which--path
information (and its relation to the conception of 
{\em complementarity}) is seen as the ultimate reason for the experimental
results, it is, in the new theory, the intrinsic information due to the
electromagnetic fields and their vector features, which are held
responsible. 

In more general terms it is the concept of the wave function and its
dogmatic interpretation, which {\em creates} the logical problems of
interpretation contained in the conventional approach:
since the wave function contains, a priori, all the information about
the particle, and since it is a scalar function, it does not contain
a vector variable. But in this case the transformation of its 
features due to a change of intrinsic directional information
must be explained by some auxiliary conception, in case of quantum erasers
by the conception of complementarity, in case of magnetic fields by the
non--local conception of spin. The whole formalism, in this
way, gains a highly artificial and abstract quality, which makes it
increasingly difficult to render a straightforward account of 
physical processes. What this analysis amounts to, is not so much
a theoretical or logical flaw of the current concept, but rather
a defect in terms theoretical simplicity.

\section{The quantum Zeno effect}\label{meas_zeno}

The quantum Zeno effect, or the unusual feature of unstable quantum
systems that measurements slow down the decay of unstable particles,
has been introduced by Misra and Sudarshan \cite{MIS77} and further 
analyzed by Peres \cite{PER89}. Its experimental implications have
recently been tested by Kwiat et al., and the theoretical feature
has been exploited to realize interaction free measurements with an
efficiency exceeding 50 \% \cite{ZEI95,KWI95}.

The proof of this feature depends on the development of an initial
state $\phi$, it shall belong to the domain of the operator $H$,
which does not explicitly depend on time. In second order 
approximation the probability of the state to survive will be, for
small time intervals $t$ ($\hbar = 1$):

\begin{equation}\label{if023}
\left|(\phi, e^{- i H t} \phi)\right|^2 \approx
1 - (\triangle H)^2 t^2 + \ldots
\end{equation}

where $t$ is assumed sufficiently small, and the finite term
$\triangle H$ shall be described by:

\begin{equation}\label{if024}
(\triangle H)^2 = (H \psi, H \phi) - (\psi, H \phi)^2
\end{equation} 

The decisive step now is to perform a measurement of the system
state $n$ times in the given interval, in this case the probability
to find the system always in its initial state will be:

\begin{equation}\label{if025}
P_{\phi} = \left[ 1 - (\triangle H)^2 (t/n)^2\right]^n >
1 - (\triangle H)^2 t^2
\end{equation}

Clearly the probability has increased, and the logical reason 
for it is the repetition of measurements. In the
infinite limit $n \rightarrow \infty$ the probability equals 1, which
obviously signifies that an unstable system in constant observation
does not decay.

In order to determine the physical content of the statement, we may
first consider an experimental realization by way of polarization
rotators \cite{KWI95}. Polarization rotators interspersed with 
polarizers can be used to establish, that the additional polarizers 
(equivalent to measurements of polarization) allow for transmission 
of photons, which are otherwise blocked out in the final 
polarization measurement. 

Seen from the viewpoint of classical electrodynamics, the result 
is completely understandable, since the transmitted intensity depends
on the angle of polarization. But from the viewpoint of quantum
theory the result is puzzling, because it indicates that the decay
of unstable states ''is not a property of the particle''\cite{MIS77}.
The paradox has been clarified to some extent by A. Peres, who
referred the effect to the features of the Schr\"odinger equation
\cite{PER89}. Expanding the time dependent wave function $\psi$
in an orthonormal basis $u_{i}$ and applying the time dependent
Schr\"odinger equation, we get:

\begin{eqnarray}\label{if026}
\psi &=& \sum a_{k}(t) u_{k} e^{- i E_{k}t} \nonumber \\
i \dot{a}_{j} &=& \sum V_{jk} a_{k} e^{i(E_{j} - E_{k})t}
\end{eqnarray}

Selecting one eigenfunction of the orthonormal basis to
represent $\phi = u_{0}$, it becomes clear that the change of
the survival amplitude is due to the amplitudes of the decay
products $a_{k}$ (since $V_{jj} = 0$). Although the result is
reasonably well explaining the  effect within the limits of
quantum theory, it is interesting to consider the same behavior
from the viewpoint of quantum ensembles. Considering a local
ensemble with exactly defined energy, the development describes
the propagation of a particle wave with energy $E$:

\begin{equation}\label{if027}
\psi(t) = \phi \,(0) \, e^{-i \omega t}
\end{equation}

And the probability $P_{\phi}$ after an arbitrary interval
$t$ is consequently equal to 1. The same applies for every component 
of a quantum ensemble in an external potential. 

Now let us assume, that
the ensemble is unstable, like the initial configuration used to 
calculate the spreading wave packet (section \ref{meas_spread}). And 
additionally, that this configuration was brought about by some kind of 
measurement process. Whether such a measurement process is possible, would 
require an extended discussion, for the quantum Zeno effect its  existence 
may be taken for granted. 
Using the result of Peres Eq. \ref{if026}, the change of this initial
configuration depends on the existence of interactions between different
components of the ensemble. As long as the measurement process is repeated
fast enough to stall the development of interacting ensemble components,
the initial configuration is conserved. The argument of Peres \cite{PER89},
that the interactions within an atomic nucleus are equivalent to a measurement
process, does not seem justified in this case. As the stable system state is
equal to a quantum ensemble of single particles, the unstable configuration
requires some sort of energy exchange. And since an isolated system cannot
provide this energy in infinite repetition, the argument then would lead to
postulating some sort of perpetuum mobile explicitly forbidden by the entropy
principle.

The quantum Zeno effect thus derives its validity from several sources:
(1) The existence of measurements leading to an unstable system state, and
(2) the provision of additional energy to regain the initial state in infinite
repetitions. From a physical point of view, the repetition contains an
infinity problem which makes the effect theoretically viable but physically
infeasible in isolated systems.

\section{Interaction--free measurements}\label{meas_intf}

\begin{figure}
\epsfxsize=1.0\hsize
\epsfbox{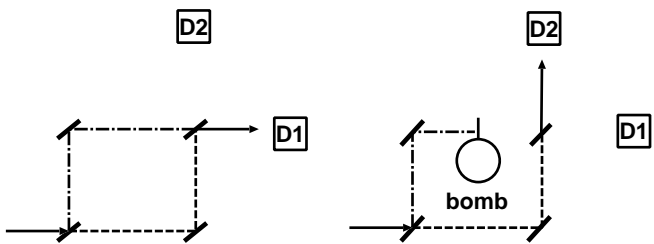}
\vspace{0.5cm}
\caption{Mach--Zehnder interferometer with or without a sensitive 
bomb--trigger in one path}\label{fig007}
\end{figure}

The essential feature of an interaction--free measurement, which
is based on a thought experiment by Renninger \cite{REN60}, is
that it provides information about the existence of an object in
a closed system without necessarily interacting with this object.
The essentials of such a measurement, recently undertaken by
Kwiat et al. \cite{ZEI95,KWI95}, can be seen in Fig. \ref{fig007}.

Interaction free measurements, usually performed with down--converted
photons, are interesting due to two features: the wave function of
the system and consequently system energy is changed, even if no
interaction occurs. And the results are seemingly incompatible with
classical field theories because the trajectories of single particles
through the measurement apparatus can be identified.

The first feature was modeled by Dicke \cite{DIC81} using a modified
Heisenberg microscope and calculating the state vectors of photons
and a non--interacting atom in first order perturbation theory. The 
result of Dicke's calculation seemed to prove that even interaction
free measurements correlate with an exchange of virtual photons or, 
in Dicke's words: {\em The apparent lack of interaction between the
atom and the electromagnetic field is only illusionary.}

On the statistical basis developed in this paper, the result must be
modified. The local modification of ensemble ranges means, in this
context, that an interaction free measurement corresponds to a 
different ensemble, i.e. an ensemble which has zero probability in
the range, where an interacting particle is appreciable. It is therefore
the limitation imposed, the change of boundary conditions, which is
the ultimate reason for the change of the wave function. And if this
local range affects system energy like in Dicke's model of an harmonic
oscillator in a magnetic field \cite{DIC81}, then the energy of the
system changes. The changed result is therefore not of physical, but
statistical origin. The logical difficulties of interpretation arise,
once more, from the Copenhagen interpretation: if wave functions in 
quantum theory are interpreted as physical determinants of single particle 
properties, interaction free measurements must indeed be referred to 
physical origins. But once this interpretation is rejected, the whole 
problem changes its qualities: instead of a physical effect it becomes a
change of statistical ensembles. And that different ensembles do have
different qualities is not all too surprising.

The second feature of interaction free measurements, the assertion by
Kwiat et al. \cite{KWI95} that ''complementary is essential'' to
the experimental results achieved, requires a critical analysis. What
the argument indicates, is the impossibility for a single photon 
in the interferometer to trigger the bomb {\em and} detector
D2 (see Fig. \ref{fig007}). Given the experimental facts, the 
argumentation is not wholly convincing. 

\begin{figure}
\epsfxsize=1.0\hsize
\epsfbox{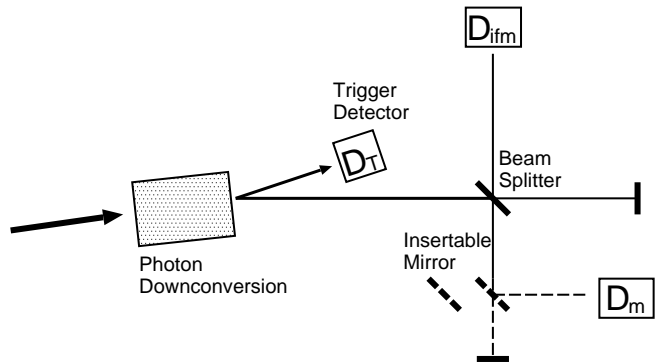}
\vspace{0.5cm}
\caption{Experiment to demonstrate the principle of interaction--free
measurements according to Kwiat et al. \protect{\cite{KWI95}}. The
insertable mirror in one arm of the Michelson interferometer equals
the bomb--in configuration, the reflectivity of the central mirror can be
changed. Note that the detector counts of detector D$_{ifm}$ and
D$_{m}$ are perfectly in accordance with classical intensities.}
\label{fig008}
\end{figure}

For their ''proof of principle'' experiment \cite{KWI95} a Michelson
interferometer was employed with an insertable mirror in one light path
(the bomb--in configuration).
The interferometer was initially adjusted in such a way, that the
detector counts on $D_{ifm}$ were a minimum, and it was shown that 
with gradual reduction of beam--splitter reflectivity the 
''figure of merit'' (the fraction of measurements allowing to
conclude the presence of the bomb without triggering it)
yields a 50 \%  probability (see Fig. \ref{fig008}). 

From the viewpoint of classical field theory the initial configuration
requires destructive interference of the two wave components in the
path of the detector. Thus the interferences, or the wave features,
are decisive in this case. If, in the bomb--in configuration, one of the
components is removed, interference is avoided and the intensity in the
detector path is the intensity of the transmitted and reflected
(TR) wave component. From the viewpoint of electromagnetic fields and
potentials, the reaction of the detector then is critical. The same
applies, basically, to the bomb detector. If detection efficiency were
high enough to allow for the conclusion that every photon is
detected at one of the two detectors, and if it could be established,
that coincidences between the two {\em detectors} certainly
do not exist, the argument of complementarity
would be valid. But as detection efficiency is only two percent, about
98 \% of the incident energy (triggering a detector by one of the
down--converted photons) is not accounted for. And in this case the
argument of complementarity as well as the whole argumentation of
interaction--free measurements seems questionable. What is missing in
the experimental results of Kwiat et al. \cite{ZEI95,KWI95} is the
proof, that triggering the detector $D_{ifm}$ is invariably related
to no energy passing the bomb detector. As long as this proof is not
provided, the results can easily be accounted for by the model of 
photons already developed \cite{HOF96B}, and which allows for 
interactions of arbitrary fractions of a single photon.

\section{Hidden variables}\label{meas_hv}

We lastly have to consider the proofs of quantum theory,
that the theory cannot contain hidden variables. A lot of
scientific research has been applied to that problem,
since the standard proofs by von Neumann \cite{NEU32} and by
Jauch and Piron \cite{JAU63} have not remained undisputed.
Especially David Bohm \cite{BOH66B} and John Bell \cite{BEL66}
criticized the proofs as inadequate. 

Bell's refutation of von Neumann's proof stresses, that 
{\it it was not the objective measurable predictions of quantum
mechanics which ruled out hidden variables. It was the arbitrary
assumption of the particular (and impossible) relation between
the results of incompatible measurements either of which might
be made on a given occasion but only one of which can in fact be
made.} His objection against the proof by Jauch and Piron stresses
the same point, i.e. that the basic assumptions contain very
peculiar properties of systems which are unnecessarily 
restrictive. An additional analysis of Gleason's proof 
\cite{GLE57} shows, in Bell's view, that this proof is in fact a circular 
argument since the implicit assumptions of the proof are essential to its
conclusion.  

The arguments of Bohm against von Neumann's and Jauch's and Piron's
proofs \cite{BOH66A,BOH66B} are similar, although centered around
the linearity requirement. 

Since the theory of measurements developed in this paper depends
explicitly on the existence of hidden variables (the intrinsic
potentials), it seems necessary to reconsider the proofs against
the very possibility of hidden variables from this angle. We shall
begin the analysis with von Neumann's proof.

\subsection{von Neumann}

Von Neumann's proof centers around an analogy with classical
statistical mechanics in his description of the hidden variables
potentially contained in quantum theory. And he stresses that the
impossibility of dispersion free states implies the impossibility
of hidden variables in quantum theory. Dispersion free states 
meaning the conception, that a quantum system is composed of a
number of individual particles with precisely determined mechanical
properties. Let us consider the mathematical procedure employed in
quantum theory for an energy measurement. The expectation value of
an operator H in a given state $\psi$ of the system is:

\begin{eqnarray}\label{hv001}
\langle H \rangle = 
\int d^3\,r\,|\psi(\vec r)|^2 H(\vec r)
\end{eqnarray}

and in case of eigenstates of H the expectation value is
equal to an precise kinetic energy $E_{k}$

\begin{eqnarray}\label{hv002}
\langle H \rangle = 
\int d^3\,r\,|\psi_{i}(\vec r)|^2 H(\vec r)
= E_{k,i}
\end{eqnarray}

Since this expectation value can be derived from Schr\"odinger's
equation, we may consider the statistical ensemble underneath the
state vector $|\psi_{i} \rangle$. First of all, the statistical
ensemble for a given energy value is a {\em homogeneous} ensemble,
meaning that it cannot be broken down in sub--ensembles with 
different statistical properties. This structure of the hidden
variables derives from their properties at every single point of
a system. For a specific eigenvalue $E_{k}$ it comprises all 
$k$--values within defined limits. And this structure of the 
statistical ensemble could only be split in components, if
different regions of the system are considered. But then the 
expectation values refer to different subsystems and are no longer
a linear combination. This requirement of linearity in the
hidden variables, on which von Neumann's proof is based, has been
disputed by Bohm \cite{BOH66B} on the grounds, that the relation
between the hidden variables and the expectation value for a
given distribution need not be linear. We will consider his
proposed theory of quantum measurements further down.

Secondly, due to the arbitrariness
in the time--variable of the Schr\"odinger equation, it is not
dispersion free, since it cannot be described as a statistical
ensemble of precisely defined physical properties. It is therefore,
in von Neumann's terminology, a {\em normal} ensemble. And as 
von Neumann only proved, that the existence of {\em non--normal}
ensembles underneath the formulations of quantum theory must be
excluded, it does not exclude the type of hidden variables found
in material wave theory. 

Bell based his refutation of von Neumann's proof on the spin
properties of particles \cite{BEL66}. As already derived, the
definition of spins in quantum theory does not allow for dispersion
free spin states of individual particles, if spin variables are
considered from the viewpoint of intrinsic variables. Therefore the
assumptions, on which von Neumann's proof is based, do not apply
also in this case.

\subsection{Jauch and Piron}

The argument of Jauch and Piron is slightly different in that it
employs projection operators into different states and 
subspaces. And they conclude that a dispersion free state defined 
by the intersection of two given states $a$ and $b$ cannot exist
\cite{JAU63}. Bell's refutation \cite{BEL66} of the argument given is 
that it refers to logical reasons not necessarily applicable to the 
statistical ensembles of hidden variables. The argument  concerning
von Neumann's proof can be repeated here. Since the states
of the intrinsic variables are not dispersion free, it cannot be
the question of referring measurements in quantum theory to an
ensemble of exactly determined particles. As soon as internal 
fluctuations due to the wave features of particles are considered,
the proof is no longer applicable.

\subsection{Bell's inequalities and spin measurements}

Bell's inequalities provided an experimental way of checking on
the existence of hidden variables. Basically, as already deduced,
the measurements of spin--variables of photons would in all likelihood
violate the uncertainty relations \cite{HOF96B}. The result of
such a measurement can only be (1) the possibility of hidden
variables is rejected, or (2) the possibility of hidden variables
is confirmed. But measurements are only valid under the condition 
that (3) the uncertainty relations have been violated. As the result 
in either case does not allow for a refutation of hidden variables 
within the framework of quantum theory, the result cannot contradict 
the theoretical background developed. 

\subsection{Bohm and Bub}

It might be interesting to consider the proposed solution of the
measurement problem in quantum theory given by Bohm and Bub 
\cite{BOH66A}. It is based on their assessment of von Neumann's proof 
and the result, that a non--linear type of relation between hidden
variables and expectation values is not explicitly excluded.

To allow for hidden variables, the authors develop a two--dimensional
subspace of state vectors $|\psi\rangle$ and $\langle \xi|$, which 
comply with a nonlinear and equally nonlocal equation of motion. 
This equation of motion determines the development of the state
vectors during measurement processes. The interesting consequence of 
the theory is, that the state vectors depend in a complicated way on
the values of the vectors in every other part of space. Their theory
allows to describe measurement processes in quantum theory due to
a fundamental irreversibility of the dynamical formulations.

We consider the argument of irreversibility in comparison with the
collapse of the wave function due to external potentials. 

On the level of individual particles an applied potential, which
shall be electrostatic, affects intrinsic variables and motion
of the particle in a deterministic way. From the viewpoint of
individual particles, irreversibility cannot be confirmed, because
electromagnetic interactions are essentially reversible. 
But considering the qualities of the ensembles before and after
the measurement, the whole process is no longer reversible, since
the full quantum ensemble cannot be obtained by an application of
a positive potential to the collapsed one. The collapse of the 
wavefunction therefore indicates an irreversersible measurement
process in quantum theory. From the viewpoint of causality it can
be understood as the selection of ensemble members with defined
qualities.

The decisive change in the measurement occurs on the level of
information. While the range of allowed $k$--values extended
from $k = 0$ to $k = k_{0}$ before the measurement, the range is
diminished after the measurement. Our information about the system
clearly has increased. Relating information to thermodynamic
entropy it could be said that a measurement process diminishes the
entropy of the system. Carrying the analogy to entropy and information one
step further, it could be said that the natural state of system should
be the state of minimum knowledge, which is, in case of 
non--interacting particles of a defined total energy, the
state where all the intrinsic potentials are equally possible. 

While, therefore, the statistical and thermodynamic qualities
do not enter in the interactions of single particles with a 
given measurement environment, they do enter on the level of 
our total knowledge about the system.  A measurement in quantum theory 
is therefore objectively defined in terms of single particle interactions and
applied external potentials. The measurement apparatus enters only
in our evaluation of the experiment, on the level of general
system entropy. Which means, that the problem of the interface
between measured objects (particles) and measurement facilities
in quantum measurements reveals an entropy principle inherent
in the framework of quantum theory by way of intrinsic 
variables and wave functions of the system.

\section{Conclusion}

We have shown in this paper that the arbitrariness of the fundamental
formulations of quantum theory signifies, theoretically, that the theory 
is to be seen as a theory of statistical manifolds. The structure of
these manifolds, called {\em quantum ensembles}, was developed for
different physical environments and the results deduced. It was also
clarified, where the logical borderline between the usually causal and
deterministic picture of classical field theory and the probabilistic
picture of quantum theory is situated, and it was established that 
classical electrodynamics gains its validity also by considering a
specific ensemble of individual wave particles, the local ensemble.
As an additional confirmation of the framework suggested, the spreading 
of the wave packet can be referred to the initial conditions commonly 
considered and which are, from the viewpoint of intrinsic potentials, 
not equilibrium states. The results are generally not, as could have 
been expected by a formal approach, independent of the physical process 
considered. The classical limit of the statistical ensembles treated 
in quantum theory was defined, it was identified as the limit of 
infinite field extension.

The question of locality was analyzed in--depth, and the results are
completely novel regarding the prevailing controversy of non--locality:
non--locality must be assumed in every formulation of fundamental
equations employing mechanical analogies. The result means, that the
Schr\"odinger equation as well as the fundamental Planck and de Broglie
relations are non--local statements. A strictly local formulation of
the theory was suggested, the Schr\"odinger equation in this case
is no longer a linear differential equation.

\begin{figure}
\epsfxsize=1.0\hsize
\epsfbox{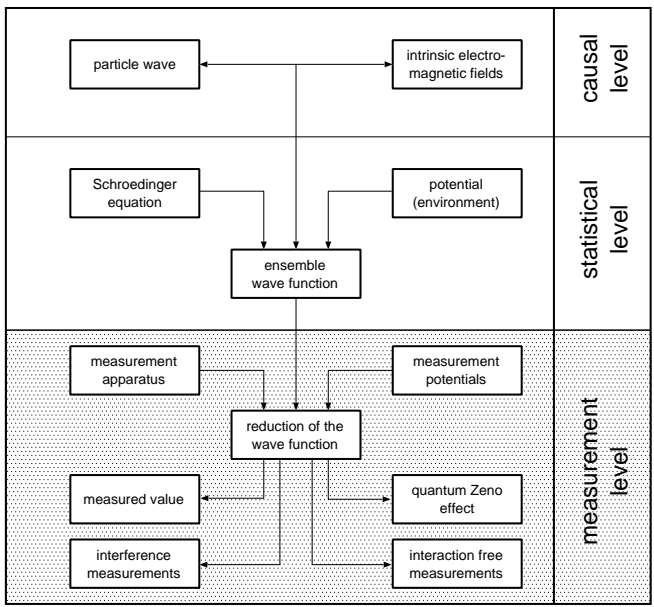}
\vspace{0.5cm}
\caption{Particles and measurements in quantum theory. The causal
level refers to individual particles, they are statistically
formalized in quantum theory as a manifold of allowed particle waves,
in measurement processes the manifold is reduced due to the
conditions of the measuring apparatus, which effects the collapse of
the wave function}\label{fig009}
\end{figure}

Fundamental measurement processes were treated in detail. The reduction
of the wave function, which currently is a major issue in attempts of
a causal reformulation of quantum theory, can be accounted for if the
change of quantum ensembles in measurement processes is considered. 
Diffraction experiments as well as neutron interference measurements
can be treated in the framework developed, the theory yields a 
{\em local} and {\em deterministic} theory of particle interactions
in magnetic fields, which does not require the non--local concept
of particle spin.

Regarding existing proofs against theories of ''hidden variables'' it
was shown that the statistical ensembles treated in quantum theory 
do not possess a structure which is explicitly excluded in the proofs
by von Neumann and Jauch and Piron. Since these ensembles cannot be
split in sub--ensembles of different physical qualities, they are, in
von Neumann's terminology {\em normal ensembles}: and the proofs
given focus specifically on the quality of non--normality. The
experimental proof by Aspect et al. \cite{ASP82}, based on Bell's
inequalities, is equally unsuitable to disprove the
statistical sublayer of quantum theory established, since the
experimental result can only be obtained, if the uncertainty 
relations are violated, as already deduced \cite{HOF96B}. A valid
violation of Bell's inequalities with regard to hidden variables
is therefore equal to a violation of fundamental principles of
quantum theory.

Generally it can be concluded, that the interface between the field theory
of particles, developed in previous papers \cite{HOF96B}, and the
actual measurement process in experimental practice can, without 
any logical inconsistency, be achieved by the suggested statistical
framework of particle waves, particle ensembles, and the change of
ensemble properties due to physical environments. Fig. \ref{fig009}
displays the logical and hierarchical structure of particle waves,
particle ensembles, and measurement processes.

\section{Discussion}

What is, therefore, it might be asked, the fundamental difference
between classical theories and quantum theory? In one or another
form this question has, since the early Twenties, raised a considerable
amount of controversy and ingenious argumentation. A part of the
problem results from an imprecise definition of the regions, where the
different theoretical frameworks apply. And the usually quoted criterion
($\hbar \longrightarrow 0$) for the classical limit is not all too
convincing, since the derivation of the classical Hamilton--Jacobi
equation from Schr\"odinger's equation is by no means unambiguous
(for a detailed account, see Holland \cite{HOL93}, chapter 6).
From a formal point of view, the difference could be interpreted
as a result of the {\em state} concept, proper to quantum theory.
But what does this concept signify apart from discrete interaction
energies: a property, which can equally be referred to energy densities
and transition rates, with the result that the energy quanta appear to
be a logical consequence of the basic mechanical outlook \cite{HOF96B}.
The differences to classical models are, furthermore, only superficial
and closely related to the fundamental interpretations employed in
quantum theory. A statement which seems to be especially appropriate
considering the logical circle of interpreting wavefunctions as
probability waves and accounting for this interpretation by a
normalization condition. From another point of view the framework
of quantum theory appears to be the kinetic counterpart of classical
electrodynamics: both theories treat, essentially, the same phenomena,
but they employ completely different methods. 

Leaving the problem of interpretation aside for the moment, the
main theoretical features of quantum theory are, in the present
context, the following:

\begin{enumerate}
\item
A micro structure of moving mass corresponding to wave like
distributions.
\item
A process of interaction described by constant transfer rates.
\item
A statistical ensemble of possible energies (quantum ensembles)
and trajectories (local ensemble).
\item
A normalization procedure which treats the manifold of single
particles as a distribution of probabilities of one single
particle.
\item
And an equation of motion for the ensembles considered 
(Schr\"odinger equation).
\end{enumerate}

It is therefore a decidedly new theoretical framework -- contrary
to the reductionist's ambitions --, and which can be understood better
by analogy. While statistical mechanics is a statistical superstructure
of classical mechanics, quantum theory is best understood as a
"statistical field theory". The result suggests to reverse the
conventional logic of development: while quantum field theory 
proceeds from foundations in quantum theory and arrives, eventually,
at the structure of electromagnetic fields, the physical qualities
of mass and charge are the foundations, on which the results in
quantum theory have to be based. Since the same should apply to the
relation between other particles and their corresponding fields,
it seems justified to consider all fundamental particles as an
expression of field type interactions and statistical measurement
processes. In itself, the approach therefore provides a new and
fairly extensive field of theoretical possibilities.

There exists, additionally, an essential difference between quantum
theory as a statistical framework and classical statistics. Statistical
mechanics is composed of single entities subject to the same physical
laws (classical mechanics), the ensemble can therefore be decomposed.
Quantum ensembles result from an imprecise and arbitrary formulation
of the fundamental law of motion (Schr\"odinger's equation). The
quantum ensemble can therefore neither be decomposed in sub--ensembles,
nor is the development of the ensemble or a possible equilibrium
state taken into account. The quantum ensemble is not calculated,
but defined by the arbitrariness of the equation of motion.

This analysis points to a gap, or rather a big hole in the 
framework of current physics, obviously brought about by the 
all too dominating Copenhagen interpretation: while a reasonable
scientific approach long ago should have questioned the foundations
of quantum physics especially in view of abundant inconsistencies
and the obvious fundamental difference between classical electrodynamics
and mechanics in their method of modelling, the prevailing mood of
mathematical indulgence served to hide the missing foundations of the 
framework developed (for a precise and up to date analysis on the 
history of this error see James Cushing \cite{CUS92}). Up to this
date the structure of ensembles, which quantum theory refers to,
has been addressed (Bohm) but not conclusively determined. The
analysis in this paper revealed that the ensembles are not determined
by physical processes -- which, in a consistent framework, they have 
to be -- but by mere {\em possibilities}, which says nothing about
the probability of members with exactly defined physical properties.
There are two solutions to this problem: (1) The proof, that the
ensembles we deal with in measurements are in fact distributions of
equal weight (in this case the current approach is justified), or
(2) the proof  that ensemble members of different properties do have
different probabilities (which is rather to be expected from 
other statistical frameworks): in this case the procedure of quantum 
theory is but a simplification. In any case it points to a missing
theoretical justification, which must be, in future, addressed.

Returning to the problem of interpretation, it can be said, that
the Copenhagen interpretation is definitely unjustified: firstly, because
the wave properties are of physical significance, and secondly, because
it does not provide a full account of physical processes. But the
same is true for the alternative interpretations: de Broglie's
interpretation is unsuitable, because the double solution is not 
correctly describing the physical events and variables (the energy
is distributed within the whole region of wave properties, although
the wave properties of the wave function are not, generally, the
wave properties of single particle waves). Bohm's interpretation
can also not be considered accurate, because the Schr\"odinger
equation is not a deterministic equation, but an equation of 
statistical ensembles. The correct interpretation is simultaneously
more complex and more simple than these approaches signify: quantum
theory has to be understood as a statistical framework of particle 
properties (consistent with the Copenhagen interpretation), although
there exists a sublayer of still more fundamental information
(consistent with the hidden variables concept): and this sublayer treats
the dynamical and intrinsic qualities of mass in motion (consistent
with de Broglie's approach), although not in a way that suggests
a double solution.

In view of the fundamental relations described by classical
electrodynamics a causal interpretation requires a {\em local}
and {\em intensive} framework of quantum theory yet to be 
formulated. The modification of Schr\"odinger's equation, derived
in section \ref{ar_nl} has not yet been tested. But the relation is
non--linear, which precludes superposition, and the potential applied
can be made consistent with the process of electron photon interaction. 

As quantum theory provides an abstract and widely applicable 
framework, the second line of research will have to establish
differences between different physical situations. It can be assumed,
that not all situations described by current quantum theory do
really require a modification of classical electrodynamics other
than including the kinetic effects (which was, for example, proved
for a simple case of Compton scattering).

Due to the presented combination of classical field theory and intrinsic
particle properties it can be expected that the picture of
micro physics eventually emerging may be far less prone to excuse
sloppy physical theories with limited human understanding. Or, in
other words, that a future theory of micro physics may be understandable
even from the viewpoint of common sense.



\end{document}